\documentclass[12pt]{article}
\usepackage{graphicx,bm,amsmath,amssymb,latexsym,epsfig,subfigure,euscript, multirow,color,verbatim}
\usepackage{bbm,pstricks}

\allowdisplaybreaks

\newcommand{\ket}[1]{\left| #1 \right>} % for Dirac bras
\newcommand{\bra}[1]{\left< #1 \right|} % for Dirac kets
\newcommand{\e}{\varepsilon}
\newcommand{\rnu}{ {\red \nu} }
\newcommand{\reta}{ {\red \eta} }

\def\tmmathbf{\mathbf}

\newcommand{\nn}{\nonumber}

\def\rd{\mathrm{d}}

\def\ecm{E_{\mathrm{CM}}}

\def\cdt{ \hspace{-0.1em} \cdot \hspace{-0.1em} }

\def\vb{\bar{v}}

\def\rd{\mathrm{d}}

\definecolor{darkred}{rgb}{0.7,0.0,0.0}
\definecolor{darkblue}{rgb}{0.0,0.0,0.9}
\definecolor{darkgreen}{rgb}{0.0,0.5,0.0}
\definecolor{brown}{rgb}{0.0,0.0,0.0}
\renewcommand{\red}{\color{darkred}}
\renewcommand{\blue}{\color{darkblue}}
\renewcommand{\green}{\color{darkgreen}}

\newcommand{\bI}{ {\blue I} }
\newcommand{\bJ}{ {\blue J} }
\newcommand{\bK}{ {\blue K} }
\newcommand{\bL}{ {\blue L} }
\newcommand{\ba}{ {\blue a} }
\newcommand{\bb}{ {\blue b} }
\newcommand{\bc}{ {\blue c} }
\newcommand{\bd}{ {\blue d} }
\newcommand{\bi}{ {\blue i} }
\newcommand{\bj}{ {\blue j} }
\newcommand{\MIJ}{  M_{\bI \bJ} }

\newcommand{\bO}{ {{\blue 1}} }
\newcommand{\bT}{ {{\blue 2}} }

\newcommand{\bpm}{ {{\blue \pm}} }

\newcommand{\DIJ}{ D_{\blue IJ} }

\newcommand{\gG}{ {\green{\Gamma} }}

\newcommand{\gp}{ {\green{+} }} 
\newcommand{\gm}{ {\green{-} }}

\newcommand{\gGG}{ {\green{\Gamma}}}

\newcommand{\LI}{ \text{Li}_2}

\newcommand{\cC}{ {\mathcal{C} }} 
\newcommand{\cA}{ {\mathcal{A} }} 
\newcommand{\cY}{ {\mathcal{Y} }}

\newcommand{\cW}{ {\mathcal{W} }} 
\newcommand{\cO}{ {\mathcal{O} }}
\newcommand{\gcusp} {\gamma_{\mathrm{cusp}}}

\newcommand{\rF}{ {{\red 4}} }
\newcommand{\rTH}{ {{\red 3}} }
\newcommand{\ri}{ {{\red i}} }
\newcommand{\rO}{ {{\red 1}} }
\newcommand{\rT}{ {{\red 2}} }
\newcommand{\ra}{ {{\red a}} }
\newcommand{\rb}{ {{\red b}} }
\newcommand{\rL}{ {{\red L}} }
\newcommand{\rR}{ {{\red R}} }
\newcommand{\rn}{ {{\red n}} }
\newcommand{\rnO}{ {{\red n_\rO}} }
\newcommand{\rnT}{ {{\red n_\rT}} }

\newcommand{\rnTH}{ {{\red n_\rTH}} }
\newcommand{\rnF}{ {{\red n_\rF}} }

\newcommand{\rofn}{{ r ( {\red n_i^\mu} ) }}

\begin{document}
\begin{titlepage}

\begin{flushright}
\end{flushright}

\vspace{0.2cm}
\begin{center}
\Large\bf
Threshold Hadronic Event Shapes \\
with Effective Field Theory
\end{center}

\vspace{0.2cm}
\begin{center}
{\sc Randall Kelley and Matthew D. Schwartz}\\
\vspace{0.4cm}
{\sl Center for the Fundamental Laws of Nature \\
Harvard University\\
Cambridge, MA 02138, USA}
\end{center}

\vspace{0.2cm}
\begin{abstract}\vspace{0.2cm}
\noindent 
Hadronic event shapes, that is, event shapes at hadron colliders, could provide a great way to test both standard and
non-standard theoretical models. However, they
are significantly more complicated than event shapes at $e^+ e^-$ colliders,
involving multiple hard directions, multiple channels and multiple color
structures. In this paper, 
hadronic event shapes are examined with
Soft-Collinear Effective Theory (SCET) by expanding around the dijet limit.
A simple event shape, threshold thrust, is defined. This observable is global and has
no free parameters, making it ideal for clarifying how resummation
of hadronic event shapes can be done in SCET. 
Threshold thrust is calculated at next-to-leading fixed order (NLO) in SCET
and resummed to next-to-next-to-leading logarithmic accuracy (NNLL).
The scale-dependent parts of the soft
function are shown to agree with what is expected from general observations, and
the factorization formula is explicitly shown to be renormalization group invariant to 1-loop.
Although threshold thrust is not itself expected to be phenomenologically
interesting, 
it can be modified into a related observable which allows the jet $p_T$ distribution
to be calculated and resummed to NNLL+NLO accuracy. As in other processes, 
one expects resummation to be important even for moderate jet momenta due to dynamical
threshold enhancement. 
A general discussion of threshold enhancement and non-global logs in hadronic event shapes is also included.
\end{abstract}
\vfil

\end{titlepage}

\section{Introduction}
Every collision at the Large Hadron Collider (LHC) will involve radiation of
quarks and gluons. In fact, the vast majority of LHC collisions will involve
nothing else (except a few photons), until the quarks and gluons hadronize and decay. These pure QCD
events provide a critical way both to test the standard and to search for new
physics. At $e^+ e^-$ colliders, an excellent way to characterize QCD events
is with event shapes. Event shapes are observables which are both infrared
safe and simple enough that a trustworthy perturbation expansion can be
performed. They usually involve a single scale, which guarantees that only one
type of large logarithm can appear. Resumming these large logarithms can then
be used to get very accurate predictions. For example, thrust was calculated
at NNLO~\cite{GehrmannDeRidder:2007hr} then resummed to N$^3$LL accuracy~\cite{Becher:2008cf}
(based on~\cite{Fleming:2007qr,Schwartz:2007ib}) and fit for power corrections~\cite{Abbate:2010xh}
using Soft-Collinear Effective Theory (SCET). 
Together, these results provided some of the best tests of QCD, and important new physics constraints~\cite{Kaplan:2008pt}.
It is therefore natural to attempt the calculation of similar observables at
hadron colliders.

At hadron colliders, pure QCD events are significantly messier than at 
$e^+e^-$ colliders. As with $e^+e^-$, it is natural to look at event shapes which
are dominated by kinematic configurations with two outgoing jets. Even then,
there are at least five new ingredients which must be understood: 1) The
energy distributions in the incoming hadrons are non-perturbative, so one must
integrate over different initial states weighted by parton-distribution
functions. 2) The process involves four directions of large energy flow (the
two protons and the two jets) as opposed to two in $e^+ e^-$; thus,
non-trivial angles are involved. 3) The 4-parton configurations involve
multiple channels ($e.g.$ $\bar{q} q \rightarrow \bar{q} q$ and $g g
\rightarrow g g$). 4) There are multiple color structures for each channel
which mix. And 5) one cannot get by with simple global observables; at a
hadron collider one must either cut out or suppress the contribution from the
region near the beam which cannot be measured. Such restrictions can generate
non-global logarithms which complicate the resummation~\cite{Dasgupta:2001sh,Dasgupta:2002dc}.

Dealing with the complication 1), the PDFs, is now well-understood. Following
the example of Drell-Yan, we can integrate over the PDFs in an effective
theory. An important lesson learned from the Drell-Yan example is that the
matching scales in the effective theory should be chosen after the convolution
with the PDFs rather than before. This avoids issues involving the Landau pole
singularity of QCD and allows for dynamical enhancement of the partonic
threshold contribution to the Drell-Yan mass~\cite{Becher:2007ty}. Similar enhancements
have also been seen in more complicated signals, such as direct photon~\cite{Becher:2009th}
and threshold $t\bar{t}$ production~\cite{Ahrens:2010zv}.

Complication 2), multiple collinear directions, has been addressed using
traditional threshold resummation in~\cite{Laenen:1998qw}
and for the hard matching coefficients 
in SCET~\cite{Chiu:2008vv,Chiu:2009ft,Chiu:2009mg,Chiu:2009yz,Becher:2009cu,Becher:2009qa,Kelley:2010fn,Fuhrer:2010eu}.
In order to understand
how angular dependence cancels in a physical cross section in SCET, a process one
step simpler than dijets was studied in~\cite{Becher:2009th}, direct photon production. Direct
photon involves three collinear directions. The SCET factorization theorem for this
case was derived in~\cite{Becher:2009th}, which required a non-trivial cancellation of
various angular factors. $t\bar{t}$ production~\cite{Ahrens:2010zv} also has multiple directions which must
cancel, but the directions are associated with the top quark velocities in the heavy quark limit
rather than with massless partons. In both cases, the cancellation was demonstrated with explicit 1-loop
calculations. Direct photon and $t\bar{t}$ also involve PDFs and dynamical threshold
enhancement as well as elements of complication 3), multiple channels.

Complication 4), multiple color structures which mix, was studied at next-to-leading order 
using traditional threshold resummation in~\cite{Kidonakis:1998nf,Botts:1989kf}.
It was also studied in SCET  for off-shell
Green's functions in~\cite{Chiu:2008vv,Chiu:2009mg,Becher:2009cu,Becher:2009qa}
and for threshold $t\bar{t}$ production in~\cite{Ahrens:2010zv}.
The dijet case is more complicated than threshold $t\bar{t}$ because of the jet functions, the myriad channels including
some with identical particles, and the necessary phase space restrictions for the soft function.
In SCET language, the color mixing occurs only in the
evolution the hard and soft functions, with the jet function and
PDF evolution being diagonal in color space. 
The hard function evolution is universal,
and observable independent.
The 1-loop Wilson coefficients for all the $2\to 2$ partonic channels 
and their renormalization
group evolution equations to 2-loops were recently calculated in SCET in~\cite{Kelley:2010fn},
based on results of~\cite{Kunszt:1993sd, Bern:1990cu, Chiu:2008vv}.
The soft function is a cross section for emission from Eikonal Wilson lines.
These emission graphs depend critically on the observable of interest.
Therefore, there is a non-trivial check on the
consistency of the factorization theorem that color evolution
of the soft function cancels that of the hard function
and its renormalization scale dependence is compensated exactly when
the hard, jet and PDF evolution equations are combined.
A general formulation of these constraints was presented in~\cite{Kelley:2010fn},
although no explicit soft function was given. A main result of this paper is
to confirm that the general results of~\cite{Kelley:2010fn} actually hold through
explicit calculation of a specific case.

Complication 5), the observable, is a difficult one. The types of observables
that will be most interesting to compute first are probably analogs of
thrust. For event kinematics in which the final state looks like
a pair of almost massless jets, thrust $\tau$ can be calculated from
the sum of the jet masses up to higher order terms
$\tau = \frac{1}{\ecm^2}( m_{{\red J_1}}^2 + m_{{\red J_2}}^2)+\cdots$,
where $m_{{\red J_1}}$ and $m_{{\red J_2}}$ are the masses of the two jets.
Jet masses are important
because the jet mass distribution is singular at tree-level  $\rd \sigma \sim
\delta (m_{{\red J_1}}^2) \delta (m_{{\red J_2}}^2)$, and therefore very sensitive to QCD radiation.
Getting the theoretical prediction for the shape of jet mass distributions to
agree with data without Monte Carlo tuning probably
requires at least the NLO distributions and resummation beyond NLL, as was
demonstrated with thrust~\cite{Becher:2008cf,Abbate:2010xh} and heavy jet mass~\cite{Chien:2010kc}
in the simpler $e^+e^-$ case.

To calculate jet masses, or a generalization of thrust, at hadron colliders
one must deal with the complication that most of the energy in a typical event
disappears down the beam pipe and is unmeasurable. Simply cutting out the beam
is unlikely to work, as it will generate non-global logarithms which cannot be
resummed. There are many ways to proceed. In a comprehensive study of hadronic
event shapes~\cite{Banfi:2010xy} with the program {\sc caesar}~\cite{Banfi:2004yd},
a number of suggestions were put forward. For
example, global transverse thrust~\cite{Banfi:2004nk} uses only transverse momentum, so that particles in
the beam direction do not contribute. Some authors have suggested using beam
functions~\cite{Stewart:2009yx}. Ref.~\cite{Banfi:2010xy} summarizes some of the
relevant issues about observables, non-global logs and the experimental measurements.

In this paper, we will take a somewhat different approach, inspired by the
practicality of dynamical threshold enhancement for the cases of Drell-Yan
and direct photon. In these cases, one calculates an
observable in the {\it machine threshold} limit: the protons collide into the final
states of interest (a lepton pair for Drell-Yan, or a photon and a jet for
direct photon) plus only soft radiation. That is, there is no outgoing
collinear field in the direction of the beam. Although the physical regime of
interest looks nothing like this -- there is a beam with a lot of energy --
resummation still seems to be quantitatively extremely useful in the physical
regime when extrapolated away from the machine threshold. This phenomenological
observation is partially understood: the logarithms of the partonic threshold
variable which should be small well away from the machine threshold are
enhanced by a factor related to the die-off of the PDFs 
near $x \rightarrow 1$~\cite{Becher:2007ty}.
To be clear, we will not attempt to demonstrate threshold enhancement for
hadronic event shapes here. 
Instead, we simply observe that the threshold region is physically motivated,
and use it to clarify some of the relevant issues in SCET.

With these motivations, we will begin in this paper by studying an observable
we call threshold thrust, which is
simply the hadronic generalization of thrust in $e^+ e^-$ events.
It can be defined exactly as regular thrust, but the factorization theorem will only
apply near the machine threshold, where the protons annihilate into two jets and soft
radiation only.
The threshold thrust distribution we define and calculate in this paper is observable, and should agree
with experiment near threshold. Unfortunately, there will not be any data
anywhere close to the machine threshold, due to the die off of the PDFs near $x
\rightarrow 1$. Thus, the observable will need to be modified so that the
large logarithms we resum will still dominate in the physical regime. 
Nevertheless, threshold thrust demonstrates a number of new features
relevant for any hadronic event shape and lets us check the general
expression for the color mixing RGE presented in~\cite{Kelley:2010fn}. Since
threshold thrust has no parameters, such as a jet size $R$,
it can be thought of as an immaculate toy model, perfect
for studying factorization in a purely hadronic environment.

After establishing and checking that threshold thrust has the expected properties,
we propose a related observable, asymmetric thrust,
which is more likely to be important phenomenologically.
Near threshold, threshold thrust can be computed, up to power corrections, by separating the event into two hemispheres
and adding the hemisphere masses.
Asymmetric thrust is defined by, instead of taking two equally-sized hemisphere jets,
we take one jet  to have size $R$, and the other jet to be everything else in the event.
Threshold thrust is not expected to be useful away from threshold because when
the beam remnants have large energy, their contribution will significantly affect hemisphere
masses. For asymmetric thrust, the beam remnants both contribute to the same jet, and so their contributions
largely cancel. Asymmetric thrust also satisfies a number of properties which we expect
observables undergoing dynamical threshold enhancement to have, as described in Section~\ref{sec:dynamic}.

The organization of this paper is as follows. In Section~\ref{sec:general}, 
we set up the calculation of hadronic events shapes in SCET. A number of features are
universal and apply to any hadronic dijet event shape. These universal features include
the 1-loop Wilson coefficients for the hard function their renormalization-group evolution equations.
We also summarize known results on inclusive jet functions and PDFs in SCET language,
since they apply to many observables. Finally, the parts of the soft function RGE 
which are off-diagonal in color space are universal, since they must cancel the hard evolution,
so we include a discussion of these as well.
Next, we introduce threshold thrust,
in Section~\ref{sec:tt}. Section~\ref{sec:soft} calculates explicitly the 1-loop soft function
for this observable. While soft function integrals are observable dependent, the color factors for emission
from an Eikonal line are universal. The way the color factors combine with the momentum-integrals
to reproduce the color structures in the hard function is non-trivial. So, this calculation 
demonstrates the kinds of calculations we expect to occur in other hadronic event shapes.
In Section~\ref{ttfinal}, we combine the ingredients into a closed-form expression for
threshold thrust which can be evaluated numerically.
In Section~\ref{sec:dynamic} we discuss how to go away from the machine threshold. We review previous examples
and catalog some lessons learned, extracting some general principles for threshold enhancement.
This leads to the proposal of asymmetric thrust, which is only briefly mentioned, and will be the subject of
future work.
Conclusions are in Section~\ref{sec:conc}.

\section{Hadronic event shapes in SCET \label{sec:general}}
In this section, we describe elements of hadronic event shape calculations which are universal,
postponing until Section~\ref{sec:tt} an actual event shape definition and event-shape specific
results. This section is essentially a summary of results derived in~\cite{Kelley:2010fn}
and can be skipped if the reader is already familiar with the contents of that paper.

In Soft-Collinear Effective Theory, a cross section is calculated by expanding around a threshold limit.  We define
{\it hadronic dijet event shapes} as observables which force that, at the exact threshold,
the incoming and outgoing partons have massless 4-momenta and that there is 
no phase space left for soft radiation. At threshold, only massless $2\to2$ scattering in QCD contributes, 
which includes the $qq \to qq$, $qq \to gg$ and $gg \to gg$ channels and their various crossings.
Going slightly away from threshold allows the partons to become
jets and for soft radiation to be produced. When the jets have
 masses much smaller than their energies $m_J \ll E_J$ and the soft radiation has energy of order
$E_{\text{soft}} \sim {m_J^2}/{E_J}$, the cross section factorizes into a convolution of
different components which are calculable separately in the effective theory.
For these event shapes, the cross section will factorize into a form 
which generically looks like
%---------------------------------------------------------------------------------------
\begin{equation} \label{genform}
 \rd \sigma \sim \rd\Pi  \sum_{\overset{\bI, \bJ}{\mathrm{channels}}}
\frac{1}{N_{\text{init}}}
H_{\bI \bJ} S_{\bJ \bI} \otimes {\mathcal J} \otimes {\mathcal J}
 \otimes f \otimes f .
\end{equation}
%-------------------------------------------------------------------------
Here, $H_{\bI \bJ}$ is the hard function, $S_{\bI \bJ}$ is a soft function, and ${\mathcal J}$ and $f$ are, respectively, jet functions
and parton distribution functions (PDFs). $\rd\Pi$ is the differential phase space and $N_{\text{init}}$ reminds us to
average over initial states. 
 $\bI$ and $\bJ$ are color indices which only
affect the hard and soft functions, since the jet functions and PDFs are color diagonal.
All the functions in this equation have an implicit channel index, which we suppress for clarity.
In the threshold limit, one can show using SCET that the channels do not interfere and the cross section for each
channel can simply be added together~\cite{Becher:2009th}.

\subsection{Wilson coefficients and the hard function}
The first step in an SCET calculation is to match QCD to SCET. This means
calculating Wilson coefficients
for SCET operators 
so that amplitudes in QCD
are reproduced at the scale $\mu$ order-by-order in perturbation theory. 
For example, one of the SCET operators relevant for $g g \to q \bar{q}$ is
%---------------------------------------------------------------------------------------
\begin{equation} \label{qqggOs}
\mathcal{O}^{stu}_{\bT \gp \gm}
 = \left(
 \bar{\chi}_{\rTH}^{\bi} {\cal A}_{\rO \perp}^{\gp \ba}
 {\green P_L} 
 {\cal A}_{\rT \perp}^{\gm \bb}\chi^{\bj}_{\rF} 
\right)
\left(
Y^\dagger_{\rTH} {\cal Y}^{\ba \ba'}_\rO 
{\blue \tau^{b'} \tau^{a'}} 
{\cal Y}^{\bb \bb'}_\rT Y_{\rF}
\right)^{\bi}_{\ \bj} \,.
\end{equation}
%---------------------------------------------------------------------------------------
Here, ${\green P_L} = \frac{1}{2}(1+\gamma_5)$ projects out the left-handed quarks while the 
gluon helicities are chosen explicitly to be $\gp$ and $\gm$ and so this operator describes 
the helicity subprocess $g^+ g^- \to q_L \bar{q}_L$. The part of the operator in the left
brackets comprises quark jets $\chi_{\rn}$  and gluon jets $\cA_{\rn \perp}$, 
which are collinear quarks and gluons
plus associated collinear Wilson lines. The right part of this operator comprises soft
Wilson lines  $Y_\rn$ and $\cY_\rn$, in the fundamental and adjoint representations, respectively.
 The soft part is independent of spin. $\bi,\bj$ are color
indices for the fundamental representation of $SU(3)$ and $\ba, \bb$ are adjoint color indices. 
For this $gg \to q\bar{q}$ channel, there are 3 color structures and 8 spin choices,
although many of the Wilson coefficients for these 24 operators are the same. There are
additional operators for the crossed channels, such as $q g \to q g$, whose Wilson
coefficients can be computed from those of $gg \to q\bar{q}$ via crossing relations. For $q q' \to q q'$ and its crossings 
there are 2 color structures, and for $gg\to gg$ there are 8 color structures. Special care is required
in computing the Wilson coefficients for $qq'\to qq'$ when $q$ and $q'$ are identical particles.
All the channels and crossings are cataloged and their Wilson coefficients computed to 1-loop in Ref.~\cite{Kelley:2010fn}.

The RG evolution of the Wilson coefficients is known to 2-loop order. 
It has the relatively simple form
%---------------------------------------------------------------------------------------
\begin{equation}
\label{Evolution_Hard}
   \frac{\rd}{\rd \ln \mu} \cC_{\bI}^\gGG (\mu) 
=   \left[ 
          \left( 
            \gamma_{\mathrm{cusp}} \frac{c_H}{2} \ln \frac{-t}{\mu^2} 
            + \gamma_H 
            -\frac{\beta(\alpha_s)}{\alpha_s}
          \right) \delta_{\bI \bJ} 
          + \gcusp \MIJ
    \right]  
    \cC_{\bJ}^\gGG (\mu) \,.
\end{equation}
%----------
Here, $\gcusp=\frac{\alpha_s}{\pi}+\cdots$ is the cusp anomalous dimension, $c_H$ is the hard group Casimir
\begin{equation}
  c_H = n_q C_F + n_g C_A \,,
\end{equation}
where $n_q$ is the number of quarks or antiquarks and $n_g$ is the number of gluons, including both initial and final partons 
(e.g. $c_H = 2 C_F + 2C_A$ for $gg \to q\bar{q}$).
 The hard anomalous dimension $\gamma_H$ can be calculated as a series in $\alpha_s$.
It is expressible in terms of quark and gluon anomalous dimensions as
\begin{equation}
  \gamma_H = n_q \gamma_q + n_g \gamma_g  \,,
\end{equation}
with $\gamma_q=\left(\frac{\alpha_s}{4\pi}\right)(-3C_F)+\cdots$ 
and $\gamma_g=\left(\frac{\alpha_s}{4\pi}\right)(-\beta_0)+\cdots$ known up to order $\alpha_s^3$~\cite{Becher:2009qa}.
The matrix $\MIJ$ describes the color mixing.  It depends on the channel and the scattering 
kinematics through logarithms of ratios of the parton momenta, however, it does not depend on $\alpha_s$. 
General expressions for $\MIJ$ are given in the color space formalism 
in~\cite{Becher:2009qa, Chiu:2009mg}. The color factors and crossings are worked out explicitly
in~\cite{Kelley:2010fn}. 
For example, for the  $qq^\prime \to q q^\prime$ channel, the mixing matrix is
%---------------------------------------------------------------------------------------
\begin{align}
  \MIJ &=
       \left(
       \begin{array}{cc}
        4C_F  (\ln\frac{-u}{s} + i \pi) 
        -C_A ( \ln\frac{t u}{s^2}+ 2i \pi ) &  
        \quad 2(\ln\frac{-u}{s} + i \pi)                        \\
        \frac{C_F}{C_A}(\ln\frac{-u}{s} + i \pi)          &  
        0 
       \end{array}
       \right)\,.
\label{qqMIJ}
\end{align}
%---------------------------------------------------------------------------------------
Explicit results for all $2 \to 2$ processes in QCD are presented 
in~\cite{Kelley:2010fn}.\footnote{
The mixing matrices given in~\cite{Kelley:2010fn} 
differ from previous results of Kidonakis et. al~\cite{Kidonakis:1998nf} only by convention-dependent diagonal terms
and a slightly different basis in the $gg\to q\bar{q}$ channels. In addition, while the results of~\cite{Kidonakis:1998nf}
are derived by computing the virtual contributions to the renormalization of soft Wilson lines with a collinear subtraction,
the $\MIJ$ matrices in ~\cite{Kelley:2010fn} are calculated simply from virtual graphs in full QCD. The results of~\cite{Kelley:2010fn}
also include the finite terms in the matching calculation and the generalization of the mixing to NNLO, which is necessary
for NNLL resummation.}

In SCET, the Wilson coefficients from matching to QCD can be combined into a hard function by
summing over spins. The hard function in any particular channel is defined by
\begin{equation}
  H_{\bI \bJ} = \sum_\gG  \cC^{\gG}_\bI  \cC^{\gG \star}_{\bJ}\,,
\end{equation}
%---------------------------------------------------------------------------------------
where $\gG$ labels the spins and $\bI$ and $\bJ$ label the color structures.
An important consequence of the color mixing being universal is the existence of  a natural basis in which 
the evolution of the Wilson coefficients, and therefore the hard function as well, are diagonal. In this basis,
%---------------------------------------------------------------------------------------
\begin{equation} 
  \frac{\rd}{\rd \ln \mu} H_{\bK \bK'}(\mu) = 
    \left[\gcusp\left( c_H \ln \left| \frac{t}{\mu^2} \right|  + \lambda_{\bK}+ \lambda_{\bK'}^\star \right)
          + 2\gamma_H
          -\frac{2\beta(\alpha_s)}{\alpha_s}
    \right]  H_{\bK \bK'}(\mu) \,,
\end{equation}
%---------------------------------------------------------------------------------------
where $\lambda_\bK$ are the eigenvalues of $\MIJ$, and we have used the fact that $\gamma_H$ is real. 
This equation is solved by
%---------------------------------------------------------------------------------------
\begin{multline}
  H_{\bK \bK'}(s,t,u,\mu) = \frac{\alpha_s(\mu_h)^2}{\alpha_s(\mu)^2}\exp 
     \Big[ 2c_H S (\mu_h, \mu)
            -2 A_H (\mu_h, \mu) \Big] \\
\times \exp\left[
            -A_{\Gamma} (\mu_h, \mu) 
              \left(\lambda_{\bK}(s,t,u)+ \lambda_{\bK'}^\star(s,t,u) +   c_H \ln\left| \frac{t}{\mu_h^2}\right| \right) 
     \right] H_{\bK \bK'}(s,t,u,\mu_h)   \,,
\end{multline}
%---------------------------------------------------------------------------------------
where
%---------------------------------------------------------------------------------------
\begin{align}
S (\nu, \mu) &= - \int_{\alpha_s (\nu)}^{\alpha_s (\mu)} d \alpha
\frac{\gamma_{\mathrm{cusp}} (\alpha)}{\beta (\alpha)} \int_{\alpha_s
(\nu)}^{\alpha} \frac{\rd \alpha'}{\beta (\alpha')}\, ,
~~~~~
A_{\Gamma} (\nu, \mu) = - \int_{\alpha_s (\nu)}^{\alpha_s (\mu)} d \alpha
\frac{\gamma_{\mathrm{cusp}} (\alpha)}{\beta (\alpha)}\,, \label{Adef}
\end{align}
%---------------------------------------------------------------------------------------
and $A_H (\nu, \mu)$ is the same  as $A_{\Gamma}(\nu,\mu)$ but with $\gamma_H$ replacing
$\gcusp$.
Closed-form expressions for these functions in renormalization-group improved perturbation theory can be found in~\cite{Becher:2006mr}.

\subsection{Jet functions  \label{ssec_Jets}}
Next, let us turn to the jet functions. Jet functions come from matrix elements of collinear
fields in SCET. The simplest jet function is the inclusive jet function, $J(m^2,\mu)$ which depends
only on the mass of the jet. This well-travelled jet function has had phenomenological application to $B$ decays~\cite{Bauer:2003pi}, 
deep-inelastic scattering~\cite{Manohar:2003vb},
event shapes~\cite{Becher:2008cf}, and direct photon production~\cite{Becher:2009th}. It is all we will need for the applications in this paper.
The inclusive quark jet function can be written to all orders as~\cite{Becher:2006nr}
%---------------------------------------------------------------------------------------
\begin{equation} \label{qjet}
 J_q (p^2,\mu) = \exp [- 4 C_F S (\mu_j, \mu) + 2 A_{J_q} (\mu_j, \mu)]
 \widetilde{j}_q (\partial_{\eta_{j q}}) \frac{1}{p^2} \left(
 \frac{p^2}{\mu_j^2} \right)^{\eta_{j_q}} \frac{e^{- \gamma_E \eta_{j_q}}}{\Gamma (\eta_{j_q})} \,.
\end{equation}
%---------------------------------------------------------------------------------------
The function $\widetilde{j}_q (L)$, which is the Laplace transform of the jet function,
can be expanded at the scale $\mu_j$ order-by-order in $\alpha_s$. Each order is a finite polynomial in $L$. This function is known exactly up
to 2-loops, and its complete $L$ dependence to 3-loops. For example, to order $\alpha_s$,
%---------------------------------------------------------------------------------------
\begin{equation}
\widetilde{j}_q (L) = 1 + \left( \frac{\alpha_s}{4 \pi} \right)
\left[4 C_F \frac{L^2}{2} -3C_F L + C_F \left( 7 - \frac{2 \pi^2}{3} \right) \right]  +\cdots \,.
\end{equation}
%---------------------------------------------------------------------------------------
The placeholder $\eta_{j_q}$ in Eq.~\eqref{qjet} is to be evaluated at $\eta_{j_q} = 2 C_F A_{\Gamma} (\mu_j, \mu)$
 after the derivatives are taken, with $A_\Gamma(\nu,\mu)$ given in Eq.~\eqref{Adef}. $A_{J_q}(\nu,\mu)$ is an evolution
kernel like $A_\Gamma(\nu,\mu)$ and $A_H(\nu,\mu)$ above. The gluon jet function has a similar form, with different 
coefficients~\cite{Becher:2010pd}. Details of all of these functions and an example application can be found in~\cite{Becher:2009th}.

To check the renormalization group equations for threshold thrust below, it is easier to use the RGE satisfied by the jet functions,
rather than to use its solution. The RGE, for either quark or gluon jets, is
%---------------------------------------------------------------------------------------
\begin{equation}
 \frac{\rd}{\rd \ln \mu} \widetilde{j}_i \left(Q^2, \mu \right) 
    = \left[ 
        - 2 C_{R_i} \gcusp \ln \frac{Q^2}{\mu^2} 
        - 2 \gamma^{J_i} 
      \right] \widetilde{j}_i \left(Q^2, \mu \right) \,,
\end{equation}
%---------------------------------------------------------------------------------------
where $C_{R_q} = C_F$, $C_{R_g}= C_A$ and the anomalous dimensions and ${\widetilde j}_i$ functions can be found in~\cite{Becher:2009th}.
To order $\alpha_s$,
%---------------------------------------------------------------------------------------
\begin{align}
    \gamma^{J_q} &= \left(\frac{\alpha_s}{4\pi}\right)(-3 C_F)+\cdots\\
 \gamma^{J_g} &= \left(\frac{\alpha_s}{4\pi}\right)(-\beta_0)+\cdots \,.
\end{align}
%---------------------------------------------------------------------------------------

\subsection{Parton distribution functions}
In any theoretical calculation at a hadron collider, parton distribution functions $f_{{\red i}/N}(x,\mu)$ for parton ${\red i}$ in nucleon $N$ will play some role. These PDFs are non-perturbative,
but their renormalization group evolution equations, the DGLAP equations, are perturbative and known to 3-loops.
In general, the evolution mixes all the different parton species. However, near the endpoint $x\to1$,
species mixing is suppressed and the evolution equations simplify.
Observables such as threshold thrust which are calculated in the threshold region where $x \sim 1$ allow
us to use this simplification to prove renormalization-group invariance. In practice, the full non-perturbative PDFs
with the general $x$ evolution can be used for phenomenology, with the $x \ll 1$ evolution compensated for at fixed
order by careful matching. Again, see~\cite{Becher:2009th} for more details and some quantitative results.

We define the Laplace transform of the parton distribution functions by
%---------------------------------------------------------------------------------------
\begin{eqnarray}
\label{Laplace_f}
\widetilde{f}_{{\red i}/N}\left(\zeta, \mu \right) 
&=&     \int_0^1 \rd x\, \exp\left(-\frac{1-x}{\zeta e^{\gamma_E} }\right) f_{{\red i}/N}(x, \mu) \,. 
\end{eqnarray}
%---------------------------------------------------------------------------------------
The evolution equations near $x = 1$ are then
%---------------------------------------------------------------------------------------
\begin{equation}
 \frac{\rd}{\rd \ln \mu} \widetilde{f}_{{\red i}/N} (\zeta , \mu) = 
    \left[ 2 C_{R_{\red i}} \gamma_{\mathrm{cusp}}
        \ln \zeta   + 2 \gamma^{f_{\red i}} 
    \right] \widetilde{f}_{{\red i}/N} (\zeta, \mu) \,.
\end{equation}
%---------------------------------------------------------------------------------------
Expressions for $\gamma^{f_q} = \gamma^{f_{\bar{q}}}$ and $\gamma^{J_g}$ to 3-loop order can be found 
in~\cite{Becher:2009th}. To order $\alpha_s$ the PDF anomalous dimension are simply negative of the $\alpha_s$ jet-function
anomalous dimensions 
%---------------------------------------------------------------------------------------
\begin{align}
    \gamma^{f_q} &= \left(\frac{\alpha_s}{4\pi}\right)(3 C_F)+\cdots\\
 \gamma^{f_g} &= \left(\frac{\alpha_s}{4\pi}\right)(\beta_0)+\cdots \,.
\end{align}
%---------------------------------------------------------------------------------------
However, at higher orders, the PDF and jet function anomalous dimensions are independent.

\subsection{Soft function: general observations  \label{ssec_Soft}}
The soft function in a hadronic event shape calculation depends on the definition of the event shape and
must be calculated for each observable separately. However, a number of features of these soft functions
are universal, and will apply for any observable. Because the renormalization group evolution is 
color diagonal for the jet functions and PDFs, the color-mixing terms in the soft function evolution
must exactly compensate the color-mixing terms in the hard function evolution. This was explained in
more detail in~\cite{Kelley:2010fn}. Here, we review those general results for completeness. A direct calculation of
the threshold thrust soft function is given in the next section, which may help clarify the notation introduced here.

The soft function is a matrix in color space. It is calculated from
matrix elements of $\cW_\bI$ which are time-ordered products
of the soft Wilson lines appearing the the SCET operators:
%---------------------------------------------------------------------------------------
\begin{equation}
    S_{\bI \bJ} \left( \{k\}, {\red n^{\mu}_i} \right) 
    =
    \sum_{X_s}
    \bra{0   } \mathcal{W}_\bI^\dagger \ket{X_s}
    \bra{X_s } \mathcal{W}_\bJ         \ket{0}  
    F_S (\{k\})  \,,
\end{equation}
%---------------------------------------------------------------------------------------
where the sum is over soft radiation in the final state. The function $F_S(\{k\})$ 
encodes  the dependence on various projections on the soft momenta related
to the definition of the observable. An explicit example for $F_S$ will be given when 
we define the observable.
No matter what the observable
is, the soft function can only depend on directions ${\red n_{i^{}}^{\mu}}$ 
of the various Wilson lines, and on arbitrary soft scales
relevant to the projections. Because of factorization, it cannot depend on the
energy of the jets, the hard scales $s, t, u$ or the energy fractions $x_{\ri}$ of
the PDFs.

As was shown in~\cite{Kelley:2010fn}, based on insights in~\cite{Becher:2009qa,Kidonakis:1998nf}, for the factorization theorem to hold,
the soft function must satisfy
%---------------------------------------------------------------------------------------
\begin{equation} \label{softRGE}
 \frac{\rd}{\rd \ln \mu} \widetilde{S}_{\bI \bJ} \Big(\{ Q \}, {\red n^{\mu}_i},\mu\Big) = 
  -\widetilde{S}_{\bI \bL} \Big(\{ Q \}, {\red n_i^\mu}, \mu\Big) \Gamma^S_{\bL \bJ}
  -\Gamma^{S\dagger}_{\bI \bL}\widetilde{S}^\dagger_{\bL \bJ} \Big(\{ Q \},{\red n^{\mu}_i}, \mu\Big)  \,,
\end{equation}
where
\begin{equation} \label{GSijgen}
 \Gamma^S_{\bI \bJ} = 
          \left( \gcusp c_Q \ln  \frac{\{Q\}} {\mu} 
          + \gcusp \rofn
          + \gamma_S \right) \delta_{\bI \bJ} 
          + \gcusp \MIJ ({\red n_i^\mu}) \, .
\end{equation}
%---------------------------------------------------------------------------------------
This soft function RGE has much in common with the RGE for the Wilson coefficients, Eq.~\eqref{Evolution_Hard}. In particular,
$\MIJ$ is the same and in both cases all the color mixing is proportional to $\gcusp$. 
The Casimir $c_Q$ controlling the Sudakov logs and the remainder function $r({\red n_i^\mu})$ depend
on the channel but are independent of $\alpha_s$. The soft anomalous dimension $\gamma_S$ 
depends on the observable and the channel but not on the color structure. 
For any particular observable, the factorization theorem will fix
$\gamma_S$ to be a linear combination of hard, jet, and PDF anomalous dimensions.

The general solution to the soft function RGE in Laplace space is
%---------------------------------------------------------------------------------------
\begin{multline} \label{genS}
  \widetilde{S}_{\bK\bK'}(\{Q\},{\red n_i^\mu},\mu) = \exp 
     \Big[ -2c_Q  S(\mu_s, \mu)  +2 A_S (\mu_s, \mu)\Big]\\
\times 
\exp \Big[   A_{\Gamma} (\mu_s, \mu)  
          \left(   
             \lambda_\bK 
            +\lambda_{\bK'}^\star  
            + \rofn
            + \rofn^\star
            + 2c_Q \ln  \frac{ \{Q\}}{\mu_s} 
          \right)
     \Big] 
\widetilde{S}_{\bK\bK'}(\{Q\},{\red n_i^\mu},\mu_s) \,,
\end{multline}
%%---------------------------------------------------------------------------------------
where $\lambda_\bK$ are the same eigenvalues of $\MIJ$ used for the hard function evolution. 
This can be transformed back to momentum space using the techniques described,
for example, in~\cite{Becher:2008cf,Becher:2009qa}.
%. The result for a soft function with one scale $k$ is
%%---------------------------------------------------------------------------------------
%\begin{multline} 
%  S_{\bK\bK'}(k,\mu) =
%     \exp\Big[ -2c_Q  S(\mu_s, \mu)  +2 A_S (\mu_s, \mu)
%+A_{\Gamma} (\mu_s, \mu)  
%          \left(   
%             \lambda_\bK 
%            +\lambda_{\bK'}^\star  
%            + \rofn
%            + \rofn^\star
%          \right)
%     \Big] 
%\\\times
%\widetilde{S}_{\bK\bK'}(\partial_{\eta_s},\mu_s) 
%\frac{1}{k}\left( \frac{k}{\mu_s}\right)^{\eta_s} 
%\frac{e^{- \gamma_E  \eta_{s}}}{\Gamma (\eta_{s})} \,,
%\end{multline}
%%---------------------------------------------------------------------------------------
%where $\eta_s = 2C_Q A_\Gamma(\mu_s,\mu)$. 
%Other examples of this notation can be found in~\cite{Becher:2008cf,Becher:2009qa}. 

We will now proceed to apply the general results of this section to a particular example: 
we define a hadronic event shape, work out its
factorization formula, calculate the soft function at order $\alpha_s$, 
check renormalization group invariance, and produce a closed form expression resummed to NNLL.

\section{Threshold Thrust \label{sec:tt}}
In the previous sections, we presented results applicable for resummation
of any observable which is dominated by dijet configurations. We saw that 
the color mixing terms in the renormalization group evolution 
can be diagonalized in a closed form to all orders in perturbation theory. 
Now we will proceed to work out the details
for a simple observable, threshold thrust. Some advantages of this observable are
that it involves only inclusive jet functions and that
all radiation contributes, making it manifestly free of non-global logs.

The factorization theorem for threshold thrust is a straightforward
combination of the ingredients used for direct photon factorization~\cite{Becher:2009th} 
and the $e^+ e^-$ thrust factorization theorem in SCET~\cite{Fleming:2007qr}. For simplicity, we will present only the
physical derivation rather than the technical one.

\subsection{Kinematics}
Thrust in $e^+ e^-$ is defined by
%---------------------------------------------------------------------------------------
\begin{equation}\label{eethrust}
 \tau =1- \max_{\vec{\rn}} \frac{\sum | \vec{p}_i \cdt \vec{\rn} |}{\sum | \vec{p}_i |} \,,
\end{equation}
%---------------------------------------------------------------------------------------
where the maximum is taken over directions $\vec{\rn}$. This somewhat complicated definition
obscures the fact that when an event has two final state jets, 
$\tau$ is simply the sum of the masses of the jets normalized to the machine energy,
up to power corrections of higher order in those masses. 

In the threshold limit for a hadronic collision, we have two incoming protons of momenta $P_\rO^{\mu}$
and $P_{\rT}^{\mu}$ annihilating into two jets with momentum $P_{\rL}^{\mu}$ and
$P_{\rR}^{\mu}$ and soft radiation. In the threshold limit, there are no beam remnants
and the jets are back-to-back. Since the are no beam remnants, the final state looks
just like a possible final state from $e^+e^-$, and so thrust can, in principle, be measured experimentally without
modification. For simplicity, we  define the jet momenta as the vector sum of the momenta of all
the radiation in the left and right hemispheres, respectively. The hemispheres are
defined with respect to the thrust axis, the maximal $\vec{\rn}$ in~Eq.~\eqref{eethrust},
which becomes equal to the jet directions in the threshold limit.
Which hemisphere is called $\rL$ or $\rR$ is arbitrary.
To further simplify, rather than using the definition in Eq.~\eqref{eethrust}, we will
simply define our threshold thrust variable $\tau$ as the sum of the hemisphere masses, normalized to the 
machine energy,  $\tau \equiv \frac{1}{\ecm^2}\left[P_{\rL}^2 +P_{\rR}^2\right]$.

Our labelling convention is that we will use $P^\mu_\rO$ and $P^\mu_\rT$ for the 
incoming protons and $p^\mu_\rO$ and $p^\mu_\rT$ for the incoming partons. We call the outgoing hemisphere momenta, 
$P^\mu_\rL$ and $P^\mu_\rR$, while the parton level quarks and gluons will be $p^\mu_\rTH$ and $p^\mu_\rF$. We allow
for either $p^\mu_\rTH$ or $p^\mu_\rF$ to align with either $P^\mu_\rL$ or $P^\mu_\rR$.
The hadronic Mandelstam variables are
%---------------------------------------------------------------------------------------
\begin{equation}
S = (P_\rO + P_\rT)^2, \hspace{1em}  
T = (P_\rO - P_\rR)^2 , \hspace{1em} 
U = (P_\rO - P_\rL)^2  \,,
\end{equation}
%---------------------------------------------------------------------------------------
and the partonic level Mandelstam variables are
%---------------------------------------------------------------------------------------
\begin{equation}
s = (p_\rO + p_\rT)^2, \hspace{1em}  
t = (p_\rO - p_\rTH)^2 , \hspace{1em} 
u = (p_\rO - p_\rF)^2  \,,
\end{equation}
with the usual definition of the momentum fractions
\begin{equation}
  p_\rO^\mu = x_{\rO} P_\rO^\mu, \quad
  p_\rT^\mu = x_{\rT} P_\rT^\mu,
\end{equation}
%---------------------------------------------------------------------------------------
At leading order, the partons are all massless. We can then define lightlike 4-vectors in the direction of the 4-momenta, so
%---------------------------------------------------------------------------------------
\begin{equation}
p^\mu_\rO \sim E_\rO\, {\red n_1^{\mu}},\hspace{1em}
p^\mu_\rT \sim E_\rT\, {\red n_2^{\mu}},\hspace{1em}
p^\mu_\rTH \sim E_\rTH\, {\red n_\rTH^{\mu}},\hspace{1em}
p^\mu_\rF \sim E_\rF\, {\red n_\rF^{\mu}}\,,
\end{equation}
%---------------------------------------------------------------------------------------
where $E_\rO, E_\rT, E_\rF, $ and $E_\rF$ are the energies of partons.

At threshold $E_\rO = E_\rT = E_\rTH = E_\rF = \sqrt{S}/2=\sqrt{s}/2=\ecm$, ${\red n_1} = \overline{{\red n}}_\rT$ and
${\red n_\rTH} = \overline{{\red n}}_{\rF}$. In this limit,
there is only one dimensionless Lorentz-invariant ratio on which the hard and soft functions can depend:
%--------------------------------------------------------------------------------------------------
\begin{equation} \label{nudef}
\rnu \equiv \frac{{\red n_1} \cdt {\red n_\rTH} }{ {\red n_1} \cdt {\red n_\rF}} 
=  \frac{{\red n_2} \cdt {\red n_\rF} }{ {\red n_2} \cdt {\red n_\rTH}} > 0 \,.
\end{equation}
%--------------------------------------------------------------------------------------------------
In terms of $s,t,u$,
%---------------------------------------------------------------------------------------
\begin{align} \label{nustu}
\rnu = \frac{t}{u}, \quad 1+\rnu = \frac{s}{-u} \,.
\end{align}
These relations will be necessary to check the factorization theorem.

To understand threshold thrust, it is helpful to pursue the analogy with direct photon production~\cite{Becher:2009th}.
For direct photon, the threshold observable was $M_X^2 = S + T + U$, where $M_X$ is the mass of everything-but-the-photon. 
In that case, $M_X$ can be written as a function of only the photon $p_T$ and rapidity
$y$, $M_X^2 = E_{\mathrm{CM}}^2 - 2 p_T E_{\mathrm{CM}} \cosh y$. For dijets,
 we can also look at $S +T +U$ but now we find that
%---------------------------------------------------------------------------------------
\begin{equation}
 S_4 \equiv S +T +U = P^2_\rR + P^2_\rL  \,.
\end{equation}
%---------------------------------------------------------------------------------------
So, in both cases $S+T+U$ gives the observable of interest. In the photon case, it is the photon $p_T$ and rapidity. In the dijet case, it the sum of the
hemisphere masses, which is equal to threshold thrust times the machine energy squared:
\begin{equation}
\tau \equiv \frac{S_4}{S} = \frac{1}{S}\left[S+T+U\right] \,.
\end{equation}
In fact, if we write this in terms of the transverse momentum $p_T$ and the rapidity $y$ of the jets,
we find an expression identical to direct photon:
\begin{equation}
\tau = 1-\frac{2p_T}{\sqrt{S}}\cosh y  \, .
\end{equation}
So, by calculating threshold thrust, we will produce the $p_T$ and rapidity spectrum of the jet. Note
that the two jets always have the same $p_T$ and opposite rapidities even away from threshold since they are hemisphere jets.

Since the kinematics are like those of direct photon production
we will express the final distribution in terms of 
the variables $v,w,p_T$ and $y$, which are related to $s,t,u,x_\rO$ and $x_\rT$ by~\cite{Gordon:1993qc,Becher:2009th},
\begin{equation} \label{vwrels}
s=\frac{1}{w}\frac{p_T^2}{v\vb}, \quad 
t=-\frac{p_T^2}{wv},\quad
u=-\frac{p_T^2}{\vb},\quad
p_T^2 = \frac{t u}{s}
\\ 
\end{equation}
\begin{equation}
x_\rO = \frac{\:p_T}{\sqrt{S}}\frac{1}{wv} e^y, \quad
x_\rT = \frac{\:p_T}{\sqrt{S}}\frac{1}{\vb} e^{-y} \,,
\end{equation}
with $\vb \equiv 1-v$.
We will use these definitions implicitly in the following. 
%The observable $\tau$ is expressible in terms of
%the $p_T$ and rapidity $y$ of the hemisphere jets by
%\begin{equation}
%  \tau = \frac{S+T+U}{S} = 1-2 \frac{p_T}{\sqrt{S}} \cosh y \,.
%\end{equation}
%So that we can equivalently measure the hemisphere $p_T$ and $y$.

To write down the threshold thrust distribution in SCET, we now use the fact that threshold thrust
is just $e^+e^-$ thrust integrated over appropriate parton luminosities with colored initial states. So we define
the partonic thrust variable
\begin{equation}
s_4\equiv m_{\rR}^2 + m_{\rL}^2  = s+t+u = \frac{p_T^2}{\vb}\frac{1-w}{w} \,,
\end{equation}
This is equivalent to the partonic variable $m_X^2$ used for direct photon in~\cite{Becher:2009th}.
Then we just write down the $e^+e^-$ thrust distribution, convoluted with PDFs:\footnote{
The $\frac{1}{w^2}$ factor in the integrand is a convention for power corrections. It is chosen so that if the integrand
were written as $\rd s_4$ instead of $\rd w$, all the nonsingular $s_4$ dependence is in the PDFs. This is slightly
different from the convention chosen in~\cite{Becher:2009th} which had all the nonsignular $w$-dependence in the PDFs.}
%---------------------------------------------------------------------------------------
\begin{multline}\label{ttHJJS}
    \frac{\rd \sigma}{\rd p_T \rd y} = \frac{1 }{8 \pi p_T}
\sum_{\text{channels}}
\frac{1}{N_{\text{init}}} 
\int_{\frac{p_T}{\sqrt{S}}e^y}^{1-\frac{p_T}{\sqrt{S}}e^{-y}} \rd v
\int_{\frac{p_T}{\sqrt{S}}\frac{1}{v}e^y}^1 \rd w \frac{v}{w^2}
        \left[x_\rO f_{\rO/N_1} (x_\rO,\mu)\right]
        \left[x_\rT f_{\rT/N_2} (x_\rT,\mu)\right]\\
\times
\sum_{\bK, \bK'}
  \frac{\alpha_s(\mu_h)^2}{\alpha_s(\mu)^2} 
H_{\bK \bK'}(v,\mu)\\
\times  \int \rd m_\rTH^2 \rd m_\rF^2 \rd k J(m_{\rTH}^2,\mu)J(m_{\rF}^2,\mu) S_{\bK \bK'}(k,v,\mu)\delta(s_4-m_{\rTH}^2 - m_{\rF}^2 - \sqrt{s} k)
\end{multline}
In contrast to $e^+e^-$ thrust, the hard and soft functions now have color indices, because the initial states are colored.
They also can depend on $s,t$ and $u$. In the threshold limit, where the hard and soft functions are calculated,
there is only one independent dimensionless ratio of these quantities, which we have taken to be $v$.

\subsection{Threshold limit}
The threshold thrust distribution calculated with SCET is only formally valid as $\tau \to 0$, equivalently, as $s\to S$. It is
in this limit that the hard and soft functions are calculated and the factorization formula can be checked. Studying this limit will
also clarify the definition of the soft function and the way the different hard directions interact.

\begin{figure}[t]
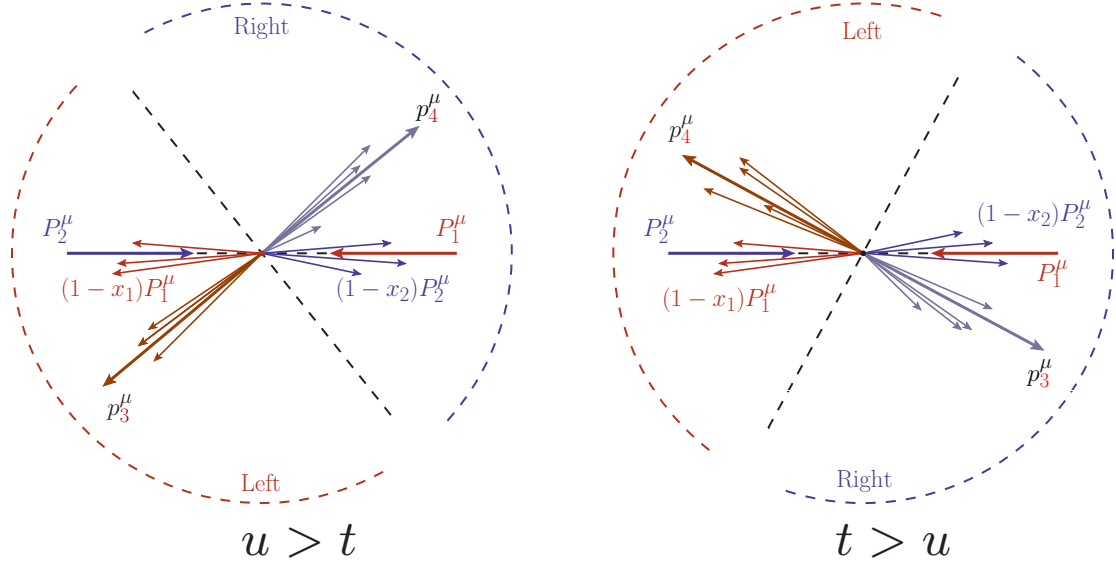

\begin{center}
%\psfrag{ugt}{$u>t$}
%\psfrag{tgu}{$t>u$}
%\psfrag{Left}{ {\red Left}}
%\psfrag{Right}{ {\blue Right}}
\includegraphics[width=0.40\hsize]{hemisphere_mass_1.epsi}
\hspace{1cm}
\includegraphics[width=0.40\hsize]{hemisphere_mass_2.epsi}
\end{center}
%\vspace{2cm}
\caption{The hemispheres are defined so that the remnants of proton $\rO$ go left and the remnants of proton $\rT$ to right. For $u>t$, the $\rTH$-jet is in the right hemisphere and the $\rF$-jet in the left hemisphere,
for $t<u$ the jets switch sides.}
\label{fig:ttkin}
\end{figure}

To derive the factorization formula, we need to relate the physical
quantities, defined at the hadron level, to perturbative quantities, defined
at the parton level. 
%As usual, we define the momentum fractions $x_\rO$ and $x_\rT$ so that
%$p^\mu_\rO = x_\rO P^\mu_\rO$ and
%$p^\mu_\rT = x_\rO P^\mu_\rT$.
The momenta $(1 - x_\rO)
P^\mu_\rO$ and $(1 - x_\rT) P^\mu_\rT$ represent the momenta from the two protons which are not
involved in the hard interaction.
In the threshold limit, this initial-state radiation is soft. 
We can then define left and right so that
that  the remnants of the first proton, $(1 - x_\rO) P^\mu_\rO$ go into the left hemisphere and
the  remnants of the second proton,
$(1 - x_\rT) P^\mu_\rT$ go into the right hemisphere.\footnote{
We could instead have defined left and right so left is aligned with parton {\red 3} and right with parton {\red 4}. However,
associating left and right with the proton directions is more obviously physical -- it does not require
the experiment to distinguish which parton is which. The threshold thrust distribution
is the same either way.
}
We also know that in the threshold limit, the hemisphere momentum scales like a collinear field and looks like a jet, {\it i.e.}
its mass is much smaller than its energy $\sqrt{P^2_\rR}
\ll E_\rR$.
So, each hemisphere must contain initial-state radiation from 
one of the proton remnants and final-state radiation from one of the jets.
This is shown graphically in Figure~\ref{fig:ttkin}. We see that there are two cases, when $u>t$, the left hemisphere aligns
with the $p_\rTH^\mu$ direction and when $t>u$ the left hemisphere aligns with the $p_\rF^\mu$ direction. 
It follows that, in the threshold limit, we can decompose the total hemisphere momentum as
\begin{align}
 P_\rR^\mu &=
 ( P^{c\mu}_\rTH + k^\mu_\rTH)\Theta(t-u) +  
 ( P^{c\mu}_\rF + k^\mu_\rF)\Theta(u-t)  + (1 - x_\rT) P^{\mu}_\rT\\
 P_\rL^\mu &=  
 ( P^{c\mu}_\rF + k^\mu_\rF)\Theta(t-u) +  
 ( P^{c\mu}_\rTH + k^\mu_\rTH)\Theta(u-t) 
+(1 - x_\rO) P^{\mu}_\rO  \,.
\end{align}
The sum $P^{c\mu}_\rn +k^{\mu}_\rn$ denotes the allocation of
 final-state radiation into collinear and soft sectors. 
This separation is not well-defined and $k_\rTH^\mu$ and $k_\rF^\mu$ must be integrated over
to form a physical observable. 
%The $\Theta$ functions separate the cases when
%${\red n_\rTH^\mu}={\red n_\rL^\mu}$ and ${\red n_\rF^\mu}={\red n_\rR^\mu}$ from the case when
%${\red n_\rTH^\mu}={\red n_\rR^\mu}$ and ${\red n_\rF^\mu}={\red n_\rL^\mu}$. This is shown graphically in Figure~\ref{fig:ttkin}.

%---------------------------------------------------------------------------------------
%---------------------------------------------------------------------------------------
The right hemisphere mass is then
%---------------------------------------------------------------------------------------
\begin{align}
 P^2_\rR &=
 \Big[m_\rTH^2 + 2 E_\rTH ( {\red n_\rTH} \cdt k_\rTH) - u (1 - x_\rT) + \cdots\Big] \Theta(t-u)\\
&~~~~~~~~~~~~~+ \Big[m_\rF^2 + 2 E_\rF ( {\red n_\rF} \cdt k_\rF) - t (1 - x_\rT) + \cdots\Big] \Theta(u-t) \,.
\end{align}
%---------------------------------------------------------------------------------------
and similarity for $P^2_\rL$. 
The ellipses denote terms of higher order in the $k_{\red i}$ or $(1-x_{\red i})$.
We have written $m_\ri =\sqrt{ ( P^{c\mu}_\ri)^2}$ to remind us that these are jet masses.
Since only one component of the soft momentum in each hemisphere
contributes, we 
we will abbreviate the projections with $k_\rTH = ({\red n_\rTH} \cdt k_\rTH)$ and $k_\rF = ({\red n_\rF} \cdt k_\rF)$. 
Also, in the partonic center-of-mass frame, the energies of the jets are $2 E_\rTH = 2 E_\rF = \sqrt{s}$.
Putting all the ingredients together, we get a cross section
of the general form of Eq.~\eqref{genform}:
%---------------------------------------------------------------------------------------
\begin{multline}
\frac{\rd \sigma}{\rd P^2_\rL \rd P^2_\rR} 
\sim
   \sum_{\overset{\bI, \bJ, \gG,}{\mathrm{channels}}} \frac{1}{N_{\text{init}}}
\int
\rd x_\rO \rd x_\rT \rd m^2_\rTH \rd m^2_\rF \rd k_\rTH \rd k_\rF
\\ \times
 ( \cC_{\bJ}^{\gG \star} \cdt S_{\bJ \bI} (k_\rTH,k_\rF) \cdt \cC_{\bI}^{\gG} )
   J_{\rTH}(m^2_\rTH) J_\rF(m^2_\rF) f_\rO(x_\rO) f_\rT (x_\rT) \\
\times\, \delta 
   \left( 
 \Big[m_\rTH^2 + \sqrt{s} ( {\red n_\rTH} \cdt k_\rTH) - u (1 - x_\rT)\Big] \Theta(t-u)
+ \Big[m_\rF^2 + \sqrt{s}( {\red n_\rF} \cdt k_\rF) - t (1 - x_\rT)\Big] \Theta(u-t)
        -P_\rR^2 
   \right)\\
 \times\, \delta \left( 
\Big[m_\rF^2 +\sqrt{s}   ( {\red n_\rF} \cdt k_\rF) - u (1 - x_\rO)\Big] \Theta(t-u)
+ 
 \Big[m_\rTH^2 +\sqrt{s}   ( {\red n_\rTH} \cdt k_\rTH) - t (1 - x_\rO)\Big] \Theta(u-t)
        -P_\rL^2 
   \right) \,.
 \label{eqwithds}
\end{multline}
%---------------------------------------------------------------------------------------
Here, the sum is over color structures $\bI$ and $\bJ$ and spins $\gG$. $\cC^{\gG}_\bI$ are the Wilson
coefficients for matching to SCET, determined by virtual graphs in full QCD.
$J_\ri(m^2)$ are quark or gluon jet functions, depending on the process, and
$f_{\red i}(x_{\red i})$ are quark or gluon PDFs. These objects were all introduced in Section~\ref{sec:general}.

Finally, threshold thrust is the sum of the hemisphere masses. So,
%---------------------------------------------------------------------------------------
\begin{align} \label{ttform}
    \frac{\rd \sigma}{\rd \tau} 
& \sim
\int \rd P_\rR^2 \rd P_\rL^2 
        \left( 
            \frac{\rd^2\sigma}{\rd P_\rR^2 \rd P_\rL^2} 
        \right)
        \delta \left(\tau -\frac{P_\rR^2 + P_\rL^2}{S}\right) \nn \\
&=
 \sum_{\overset{\bI, \bJ}{\text{channels}}} \frac{1}{N_{\text{init}}}
\int
\rd x_\rO \rd x_\rT \rd m^2_\rTH \rd m^2_\rF \rd k  
\\&~~~~~~~\times 
H_{\bI \bJ}  S_{\bJ \bI} ( k )
   J_{\rTH}(m^2_\rTH) J_{\rF}(m^2_\rF)  f_\rO(x_\rO) f_\rT (x_\rT)  \nn\\
&~~~~~~~~~~~~~~
\times
    \delta 
    \left(
      m_\rTH^2 + m_\rF^2 
      +  \sqrt{s} k
      -\min(t,u) (1 -x_\rO) - \min(t,u)(1 - x_\rT) 
      -S \tau 
      \right) \nn  \,,
\end{align}
%---------------------------------------------------------------------------------------
where the threshold thrust soft function is defined as
\begin{align}
  S_{\bJ \bI}^T(k) &\equiv \int
 \rd k_\rTH \rd k_\rF  S_{\bJ \bI} (k_\rTH,k_\rF)
\delta( k - k_\rTH - k_\rF) \,.
\end{align}
We will now show how to calculate all of the objects in this formula and check
the renormalization scale independence to order $\alpha_s$.

\subsection{Soft function \label{sec:soft}}
Now that we have defined an observable, threshold thrust, which
is the sum of the hemisphere masses in the threshold limit,
we can define the soft function more precisely. In this case, it
%---------------------------------------------------------------------------------------
\begin{equation}
 \label{S_def}
 S_{\bI \bJ} (k_\rTH, k_\rF) = 
  \sum_{X_s} \bra{0} \mathcal{W}^\dagger_\bI \ket{ X_s}\bra{ X_s } \mathcal{W}_\bJ \ket{0} 
  \delta \left( {\red n_\rTH} \cdt P^{X_s}_\rTH - k_\rTH \right) 
  \delta \left( {\red n_\rF} \cdt P^{X_s}_\rF - k_\rF \right) \,,
\end{equation}
%---------------------------------------------------------------------------------------
where $P^{X_s}_\rTH$ and $P^{X_s}_\rF$ are the sum of the momenta in the particles in state
$X_s$ which go into the hemispheres containing parton $\rTH$ or $\rF$ respectively.  

As an example, consider the $qq' \to qq'$ channel.
For this channel the $\cW_\bI$ are constructed out of soft 
Wilson lines  $Y_\rn$ in the fundamental representation. The two color structures are
%---------------------------------------------------------------------------------------
\begin{eqnarray} \label{qqWs}
 \cW_{\bO} &=& \mathbf{T} 
    \left\{ 
        (Y^\dagger_\rF {\blue \tau^a} Y_\rT ) 
        (Y^\dagger_\rTH {\blue \tau^a} Y_\rO)  
    \right\}  \nonumber \\
 \cW_{\bT} &=& \mathbf{T} 
    \left\{ 
        (Y^\dagger_\rF {\blue \mathbbm{1}} Y_\rT ) 
        (Y^\dagger_\rTH {\blue \mathbbm{1}} Y_\rO)  
    \right\} .
\end{eqnarray}
%---------------------------------------------------------------------------------------
Here, $\mathbf{T}\{ \}$ stands for time ordering and $\mathbbm{1}$ is the identity operator in $SU(3)$.
The $\mathcal{W}_{\bI}$ have external fundamental indices, which have been suppressed for conciseness.
These indices get contracted with each other when the $\cW_\bI$ are combined into
the gauge invariant soft function.
The more explicit expression is
%---------------------------------------------------------------------------------------
\begin{multline}
 S_{\bI \bJ} = 
  \sum_{X_s} \big\langle0\Big| \tmmathbf{\bar{T}}  
    \left\{  
      (Y_\rO^\dagger {\blue T^{\dagger}_I} Y_\rTH )^{\blue \ i_1}_{\blue i_3} 
      (Y_\rT^\dagger {\blue T^{\dagger}_I} Y_\rF )^{\blue \ i_2}_{\blue i_4} 
    \right\}  
    \Big| X_s \Big\rangle  \\
\times     \big\langle X_s\Big|  \tmmathbf{T}  
    \left\{  
      (Y_{\rF}^{\dagger} {\blue T_J} Y_{\rT} )^{\blue \ i_4}_{\blue i_2} 
      (Y_{\rTH}^{\dagger} {\blue T_J} Y_{\rO} )^{\blue \ i_3}_{\blue i_1} 
    \right\} \Big| 0 \Big\rangle \,
  \delta \left( {\red n_\rTH} \cdt P^{X_s}_\rTH - k_\rTH \right) 
  \delta \left( {\red n_\rF} \cdt P^{X_s}_\rF - k_\rF \right) \,,
\label{Sexplicit}
\end{multline}
%---------------------------------------------------------------------------------------
where ${\blue T_1 \otimes T_1 = \tau^a \otimes \tau^a}$ and 
${\blue T_2 \otimes T_2 = \mathbbm{1} \otimes \mathbbm{1}}$.

The hemisphere soft function is complicated by the necessary projections into the hemispheres, 
which prevents
us from simply summing over the intermediate states.
At tree level, these projections are trivial, the Wilson lines are numbers, $Y_\rn =1$,
and the soft function for $qq'\to qq'$ evaluates to
\begin{align} \label{softtree}
S^\text{tree}_{\bI \bJ}(\{ k \}, {\red n_i} ) 
&=  
\begin{pmatrix}
  \text{Tr}[{\blue \tau^b \tau^a }] \text{Tr}[{\blue \tau^a \tau^b }] & 
  \text{Tr}[{\blue \tau^a }] \text{Tr}[{\blue \tau^a} ]               \\
  \text{Tr}[{\blue \tau^b }] \text{Tr}[{\blue \tau^b} ]               & 
  \text{Tr}[{\blue \mathbbm{1} }] \text{Tr}[{\blue \mathbbm{1} }]     
\end{pmatrix}
 \delta (k_\rTH) \delta (k_\rF)
\nn \\ 
&=
\begin{pmatrix}
  \frac{C_A C_F}{2} & 0 \\
  0                 & C_A^2 
\end{pmatrix}
 \delta (k_\rTH) \delta (k_\rF) \,.
\end{align}
%---------------------------------------------------------------------------------------
The color factors here are related to the normalization of the Wilson line in Eq.~\eqref{qqWs}. 
Careful monitoring of the normalization is required for the factorization theorem to check.

\subsubsection{NLO soft function}
\begin{figure}[t]
\begin{center}
\includegraphics[width=0.95\hsize]{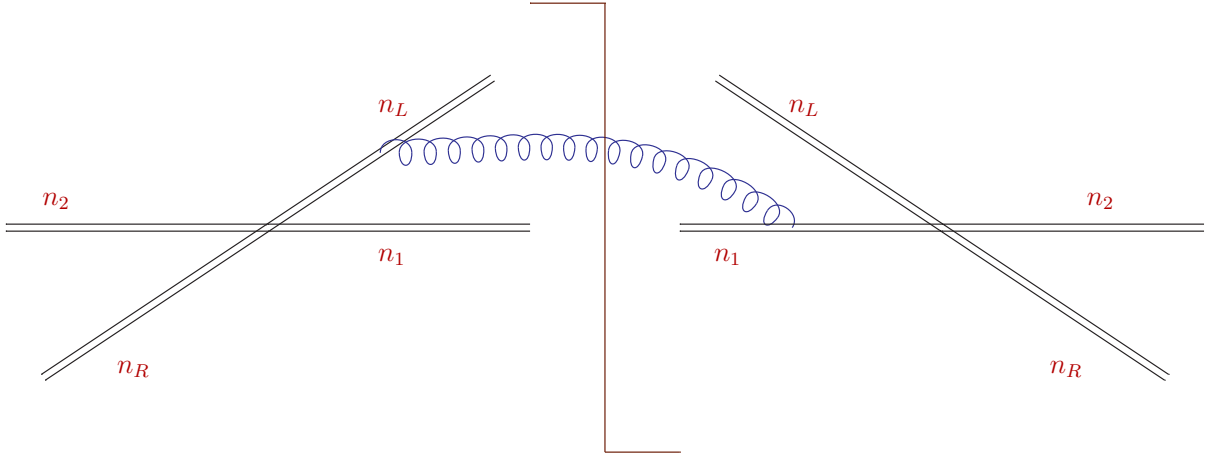}
\end{center}
\caption{Diagrams contributing to the soft function at NLO.}
\label{fig:soft}
\end{figure}

At order $\alpha_s$, there are contributions to $S_{\bI \bJ}$ from virtual and real emission graphs, 
however. in dimensional regularization, the virtual graphs for matrix elements involving
only soft Wilson lines are scaleless and vanish. The real emission contribution
involves emissions from any of the Wilson lines and absorption into any other.
The necessary diagrams can be drawn as cuts, as shown in Figure~\ref{fig:soft}.
The calculation can be split into two parts: calculation of the
integrals, which we call $I_S$, and calculation of the color factors, which we call $D_{\bI \bJ}$. 
The $\cO(\alpha_s)$ result can then be written as
%---------------------------------------------------------------------------------------
\begin{equation} 
\label{schan}
S_{\bI \bJ}^{\text{NLO}}(k_\rTH,k_\rF) 
=
 \sum_{\ra, \rb}
  I_S( {\red n_a}, {\red n_b},k_\rTH,k_\rF ) 
  D_{\bI \bJ}({\red a}, {\red b}) \,.
\end{equation}
%---------------------------------------------------------------------------------------
The sum is over assignments of $\ra$ and $\rb$ to any of the four Wilson lines $\rO, \rT, \rTH, \rF$.
The color factors are matrices in color space and depend on the channel being considered. 
The integrals depend only on kinematic factors not depend on the color nor channel, {\it i.e.} 
whether the lines are quarks, anti-quarks or gluon. A similar computation was done for $t\bar{t}$ production
in~\cite{Ahrens:2010zv}.

%---------------------------------------------------------------------------------------
%---------------------------------------------------------------------------------------

%---------------------------------------------------------------------------------------

The integral for emission from leg ${\red n_a^\mu}$ leg and absorption into the ${\red n_b^\mu}$ leg,
in $d=4-2\e$ dimensions, is
%---------------------------------------------------------------------------------------
\begin{align}
& I_S ( {\red n_a}, {\red n_b}, k_\rTH, k_\rF) =  
g_s^2
 \left( \frac{ \mu^{2} e^{\gamma_E} }{4\pi} \right)^\e 
    \int \frac{d^d q}{(2\pi)^{d-1} } \Theta (q^0) \delta (q^2) 
    \frac{{\red n_a} \cdt {\red n_b}}{( {\red n_a} \cdt q) ( {\red n_b} \cdt q)}
    \nonumber \\
& \quad \times  
    \Bigl\{
        \Theta ( {\red n_\rTH} \cdt q - {\red n_\rF} \cdt q)
        \delta (k_\rF -{\red n_\rF} \cdt q) 
        \delta (k_\rTH ) 
        + 
        \Theta ({\red n_\rF} \cdt q - {\red n_\rTH} \cdt q) 
        \delta (k_\rTH - {\red n_\rTH} \cdt q) 
        \delta (k_\rF ) 
    \Bigr \} \, .
\end{align}
%---------------------------------------------------------------------------------------
The second line denotes the projections into the two hemispheres: its first term says ``If the
component of $q^\mu$ going to the $\rF$ direction is larger than the component going to 
the $\rTH$ direction, then the soft radiation must be going to the $\rF$ direction. 
It therefore only contributes to $k_\rF$, by an amount equal to
${\red n_\rF} \cdot q$.'' The second term is for the case where the radiation goes in the $\rTH$
direction.  At order $\alpha_s$ the radiation can either go in the $\rTH$ or $\rF$ direction, 
but not both. 

Because of the explicit appearance of ${\red n_\rTH^\mu}$ and
${\red n_\rF^\mu}$, the integral has a different form for the various assignments of  ${\red n_a}$ and ${\red n_b}$
to the various Wilson line directions.
To simplify the expressions, we can use the fact that ${\red n_\rTH^\mu}$ and ${\red n_\rF^\mu}$ 
are back-to-back as are ${\red n_1^\mu}$ and ${\red n_2^\mu}$. 
It is also true that the integrals 
are Lorentz invariant as well as invariant under separate rescalings of ${\red n_1^\mu}$ and 
${\red n_2^\mu}$. Using these observations, the integrals can only be functions of
the single ratio $\rnu = {{\red n_1} \cdt {\red n_\rTH} }/{ {\red n_1} \cdt {\red n_\rF}}$,
as in Eq.~\eqref{nudef}.
The integrals are not smooth functions of $\rnu$ at $\rnu=1$, due to the convention chosen for the color
basis; a different convention might move the poles from the $t$- to $u$- channel in a particular color operator.
We give all the integrals for all values of $\rnu$ in Appendix~\ref{Appendix_Soft}.
Some of the cases already exist in the literature. For example, if ${\red n_a}= {\red n_\rTH}$ 
and ${\red n_b} = {\red n_\rF}$, then the integral is the same one from the 
hemisphere soft function from thrust, which
was calculated in~\cite{Schwartz:2007ib} and~\cite{Fleming:2007xt}. 
Another case can be checked against results in~\cite{Ellis:2010rw}. In all cases,
our results agree with previous work.

In addition to performing the integrals, the color factors $\DIJ({\red a}, {\red b})$ have to be worked out. 
These color factors come from traces of products of group generators
coming from the expansion of the various Wilson lines to NLO.
They depend on which line is emitting and which is absorbing, and therefore on
the assignments of $\ra$ and $\rb$.
The NLO color factors for all channels are given in Appendix~\ref{Appendix_Soft}.
The color factors depend only on which
Wilson lines are doing the emitting, not on the details of the observable (unlike the integrals).
Therefore, these
color factors should be generally useful for the calculation of any dijet hadronic event shape, 
not just threshold thrust.

%--------------------------TABLE--QUARK----------------------------------------------------------------
\begin{table}[t] 
\begin{center}
  \begin{tabular}{|c|c|c|c||c|c|c|c|} 
\hline
$\rO\rT \to \rTH \rF$ & \text{cross} &$\chi ({\red a})$ & $\rofn$ &
$\rO\rT \to \rTH \rF$ & \text{cross} &$\chi ({\red a})$ & $\rofn$\\
\hline
  \multirow{3}{*}{$
                    \begin{array}{c}
                    q q' \to q q' \\
                    (q q \to q q) \\
                    \end{array}
                  $}
& \multirow{3}{*}{$s t u$}
& \multirow{3}{*}{${\red 1 2 \rTH \rF}$}
& \multirow{3}{*}{$
                    \begin{array}{c}
                    2C_F  \ln\frac{t}{\min(t,u)} \\ 
                    \end{array}$} 
& \multirow{3}{*}{~$q q' \to q' q$}
& \multirow{3}{*}{$s u t$}
& \multirow{3}{*}{${\red 1 2 \rF \rTH}$}
& \multirow{3}{*}{$
                    \begin{array}{c}
                    2C_F  \ln\frac{u}{\min(t,u)} \\ 
                    \end{array}$} 
\\
&
&
&
&
&
&
&
\\
&
&
&
&
&
&
&
\\
\hline
  \multirow{3}{*}{$
                    \begin{array}{c}
                     q \bar{q}' \to q \bar{q}' \\
                    (q \bar{q}  \to q \bar{q})
                    \end{array}
                  $}
& \multirow{3}{*}{$u t s$}
& \multirow{3}{*}{${\red 1 \rF \rTH 2}$}
& \multirow{3}{*}{$
                    \begin{array}{c}
                    2C_F  \ln\frac{t}{\min(t,u)} \\ 
                    \end{array}$} 
& \multirow{3}{*}{~$q \bar{q}' \to \bar{q}' q$}
& \multirow{3}{*}{$t u s$}
& \multirow{3}{*}{${\red 1 \rF 2 \rTH}$} 
& \multirow{3}{*}{$
                    \begin{array}{c}
                    2C_F  \ln\frac{u}{\min(t,u)} \\ 
                    \end{array}$} 
\\
&
&
&
&
&
&
&
\\
&
&
&
&
&
&
&
\\
\hline
  \multirow{3}{*}{$
                    \begin{array}{c}
                     q \bar{q} \to \bar{q}' q' \\
                    (q \bar{q} \to \bar{q} q)
                    \end{array}
                  $}
& \multirow{3}{*}{$t s u$}
& \multirow{3}{*}{${\red 1 \rTH 2 \rF}$}
& \multirow{3}{*}{$
                   \begin{array}{c}
                    2C_F ( \ln\frac{s}{-\min(t,u)}  \\ 
                    -i\pi )
                    \end{array}$} 
&
  \multirow{3}{*}{~$q \bar{q} \to q' \bar{q}'$}
& \multirow{3}{*}{$u s t$}
& \multirow{3}{*}{${\red 1 \rTH \rF 2}$}  
& \multirow{3}{*}{$
                    \begin{array}{c}
                    2C_F ( \ln\frac{s}{-\min(t,u)}  \\ 
                    -i\pi )
                    \end{array}$} 
\\
&
&
&
&
&
&
&
\\
&
&
&
&
&
&
&
\\
\hline
  \end{tabular}
\caption
  {  \label{tab:perm1}
Crossing relations for the color factors in the soft function for the $qq \to qq$ channels. 
Identical particle channels are bracketed.
The ``cross'' column show the permutation
of Mandelstam invariants used to cross the Wilson coefficients. 
The $\chi(\ra)$ columns are the permutations
used in $\DIJ(\chi(\ra), \chi(\rb))$.
The $\rofn$ columns show the explicit crossed form of $\rofn$ appearing in
$\Gamma^S_{\bI \bJ}$, in Eq.~\eqref{softgenB}.}
\end{center}
\end{table}
%--------------------------TABLE------------------------------------------------------------------

%--------------------------TABLE----Gluon--------------------------------------------------------------
\begin{table}[t] 
\begin{center}
  \begin{tabular}{|c|c|c|c||c|c|c|c|} 
\hline
$\rO\rT \to \rTH \rF$ & \text{cross} &$\chi ({\red a})$ & $\rofn$ &
$\rO\rT \to \rTH \rF$ & \text{cross} &$\chi ({\red a})$ & $\rofn$\\
\hline
  \multirow{3}{*}{$
                    \begin{array}{c}
                    gg        \to q \bar{q} \\
                    q \bar{q} \to gg 
                    \end{array}
                  $} 
& \multirow{3}{*}{$s t u$}
& \multirow{3}{*}{$
                    \begin{array}{c}
                    {\red 1 2 3 4}\\ 
                    {\red 4 3 2 1}
                    \end{array}
                  $} 
& \multirow{3}{*}{$
                    \begin{array}{c}
                    c_m  \ln\frac{t}{\min(t,u)} \\ 
                    +\frac{c_Q}{2} \ln \frac{s}{-t}
                    \end{array}
                  $} 
&
  \multirow{3}{*}{$
                    \begin{array}{c}
                    gg        \to \bar{q} q \\
                    \bar{q} q \to gg 
                    \end{array}
                  $} 
& \multirow{3}{*}{$s u t$}
& \multirow{3}{*}{$
                    \begin{array}{c}
                    {\red 1 2 4 3}\\ 
                    {\red 3 4 2 1}
                    \end{array}
                  $} 
& \multirow{3}{*}{$
                    \begin{array}{c}
                    c_m  \ln\frac{u}{\min(t,u)} \\ 
                    +\frac{c_Q}{2}\ln \frac{s}{-u}
                    \end{array}$} 
\\
&
&
&
&
&
&
&
\\
&
&
&
&
&
&
&
\\
\hline
  \multirow{3}{*}{$
                    \begin{array}{c}
                    \bar{q} g \to g \bar{q} \\
                    g q \to q g
                    \end{array}
                  $} 
& \multirow{3}{*}{$u t s$}
& \multirow{3}{*}{$
                    \begin{array}{c}
                    {\red 3 2 1 4}\\ 
                    {\red 1 4 3 2}
                    \end{array}
                  $} 
& \multirow{3}{*}{$
                    \begin{array}{c}
                    c_m  \ln\frac{t}{\min(t,u)}
                    \end{array}
                  $} 
&
  \multirow{3}{*}{$
                    \begin{array}{c}
                    \bar{q} g \to \bar{q} g\\
                    g q \to g q 
                    \end{array}
                  $} 
& \multirow{3}{*}{$t u s$}
& \multirow{3}{*}{$
                    \begin{array}{c}
                    {\red 3 2 4 1}\\ 
                    {\red 1 4 2 3}
                    \end{array}
                  $} 
& \multirow{3}{*}{$
                    \begin{array}{c}
                    c_m  \ln\frac{u}{\min(t,u)}
                    \end{array}$} 
\\
&
&
&
&
&
&
&
\\
&
&
&
&
&
&
&
\\
\hline
  \multirow{3}{*}{$
                    \begin{array}{c}
                    q g \to q g \\
                    g \bar{q} \to g \bar{q}
                    \end{array}
                  $} 
& \multirow{3}{*}{$t s u$}
& \multirow{3}{*}{$
                    \begin{array}{c}
                    {\red 4 2 3 1}\\ 
                    {\red 1 3 2 4}
                    \end{array}
                  $} 
& \multirow{3}{*}{$
                    \begin{array}{c}
                    c_m ( \ln\frac{s}{-\min(t,u)} \\
                    -i\pi)
                    \end{array}
                  $} 
&
  \multirow{3}{*}{$
                    \begin{array}{c}
                    q g \to g q \\
                    q \bar{q} \to \bar{q} g
                    \end{array}
                  $} 
& \multirow{3}{*}{$u s t$}
& \multirow{3}{*}{$
                    \begin{array}{c}
                    {\red 4 2 1 3}\\ 
                    {\red 1 3 4 2}
                    \end{array}
                  $} 
& \multirow{3}{*}{$
                    \begin{array}{c}
                    c_m ( \ln\frac{s}{-\min(t,u)} \\
                    -i\pi)
                    \end{array}$} 
\\
&
&
&
&
&
&
&
\\
&
&
&
&
&
&
&
\\
\hline
  \end{tabular}
\caption
  {  \label{tab:perm2}
Crossing relations for the color factors in the soft function for the $gg\to q\bar{q}$ channels, with columns as in Table~\ref{tab:perm1}. $c_m$ and
$c_Q$ are given in Eqs.~\eqref{cmeq} and \eqref{cQeq}, and $c_Q$ has been dropped when it vanishes. There
are more soft functions for $gg\to q\bar{q}$ than for $qq \to qq$ since $D_{\bI \bJ} (\rO,\rT) \ne D_{\bI \bJ}(\rTH, \rF)$ in this
case.
}
\end{center}
\end{table}
%--------------------------TABLE------------------------------------------------------------------

To compute the soft function for a crossed process one must permute which
Wilson lines go into the color factors with respect to Wilson lines used
for emissions in the associated integral. To do this, we write
\begin{equation}
 D_{\bI \bJ}({\red a},{\red b}) \to D_{\bI \bJ}( {\chi( {\red a} )},{\chi({\red b})})  \,.
\label{perm}
\end{equation} 
In general, there are 4! = 24 permutations, but they are not all independent. For the $qq \to qq$ there
are 6 independent permutations $\chi({\red a})$ which are listed in Table~\ref{tab:perm1}. The other 18 channels
are equal to one of these using either charge conjugation ($q\leftrightarrow \bar{q}$ for all 4 quarks)
or $q \leftrightarrow q'$.
The permutations for the $gg \to q\bar{q}$ channels are listed in Table~\ref{tab:perm2}.
In this case, there 12 different permutations corresponding to the 12 different physical channels.
For some channels, like $q\bar{q} \to gg$, there is an ambiguity as to whether it is an $stu$ channel with 
$\chi(\ra) = {\red 4321}$ or an $sut$ channel with $\chi(\ra) = {\red 4312}$. We can pick either
convention, and as long as the appropriate $stu$ or $sut$ hard function is used, the physical cross section will be
the same. 
For $gg \to gg$ there is only one channel, so it does not get a table.

To be clear about the $\chi({\red a})$ notation, consider two examples. For the $q q' \to q q'$ channel,
corresponding to  $stu$ and $\chi(\ra) = {\red 1234}$ in Table~\ref{tab:perm1},
the soft function is
\begin{align}
  S^{{\red 1234}}_{\bI \bJ} &=2\Big[ I_S(\rnO, \rnT) \DIJ (\rO, \rT) + I_S(\rnO, \rnTH) \DIJ (\rO, \rTH) + I_S(\rnO, \rnF) \DIJ (\rO, \rF)\\
 &~~ + I_S(\rnT, \rnTH) \DIJ (\rT, \rTH) + I_S(\rnT, \rnF) \DIJ (\rT, \rF) + I_S(\rnTH, \rnF) \DIJ (\rTH, \rF)\Big] \,,\nn
\end{align}
where the $2$ comes from $D_{\bI \bJ}(\ra, \rb) = D_{\bI \bJ}(\rb, \ra)$.
For $q \bar{q} \to \bar{q}' q'$, 
which according to Table~\ref{tab:perm1}  is $tsu$ with $\chi(\ra) = {\red 1 3 2 4}$, the soft function is
\begin{align}
  S^{tsu}_{\bI \bJ} &=2\Big[ I_S(\rnO, \rnT) \DIJ (\rO, \rTH) + I_S(\rnO, \rnTH) \DIJ (\rO, \rT) + I_S(\rnO, \rnF) \DIJ (\rO, \rF)\\
 &~~ + I_S(\rnT, \rnTH) \DIJ (\rTH, \rT) + I_S(\rnT, \rnF) \DIJ (\rTH, \rF) + I_S(\rnTH, \rnF) \DIJ (\rT, \rF)\Big] \,,\nn
\end{align}
and so on for the other channels. 
In our notation, $q'$ and $q$ are different flavors of quarks (for example, up and down). The
soft function does not care if the particles are identical. However, for identical particles, one must choose a convention
for which color basis to use. Our convention agrees with the convention in~\cite{Kelley:2010fn}, and the appropriate
permutations are listed in Table~\ref{tab:perm1}.

Since the soft functions are calculated with real emissions integrals, they are all real. Thus, the crossings
cannot be computed by expressing the final soft function in terms of $stu$ and then crossing $stu$,
as we do for the Wilson coefficients. In fact, there is no obvious relationship between the soft functions
for the different crossings and they simply have to be computed separately by permuting the arguments of $\DIJ (\ra, \rb)$.

As an example, combining the integrals and color factors for the $qq' \to qq'$ channel,
 the threshold thrust soft function is
%---------------------------------------------------------------------------------------
\begin{multline}
S_{\bI \bJ}(k) = \delta(k) 
  \begin{pmatrix}
            \frac{1}{2}C_F C_A      &  0  \\
                      0             &  C_A^2
  \end{pmatrix} 
+\left(\frac{\alpha_s}{4\pi}\right)  c^S_{\bI \bJ}(\rnu)\delta(k)\\
+\left(\frac{\alpha_s}{4\pi}\right)   
\begin{pmatrix}
    4  C_F \ln\frac{(1+\rnu)^2}{\rnu} - 8 C_F^2 C_A \ln\rnu \,\Theta(1-\rnu) &
    \quad  -8C_A C_F \ln (1 + \rnu) \\
    \quad  -8C_A C_F \ln (1 + \rnu) &
    \quad   16 C_A^2 C_F \ln \rnu\, \Theta(\rnu-1)
\end{pmatrix}
\left[\frac{1}{k}\right]_\star^{ [k,\mu]} \,.
\end{multline}
%---------------------------------------------------------------------------------------
Here, $[1/k]_\star^{[k,\mu]}$ is a star distribution (see~\cite{Schwartz:2007ib}).
For the $\mu$-independent part, $c^S_{\bI \bJ}(\rnu)$, see Eq.~\eqref{csqqqq}.  Other channels will have a different $c^S_{\bI \bJ}(\rnu)$. Also, 
for other channels, there may be an additional piece proportional 
to $[\ln k/k]_\star^{[k,\mu]}$.  It just so happens that the coefficient of this additional 
term vanishes for $qq'\to qq'$ (and its crossings).
%The appearance of  $\Theta(1-\rnu)$ is expected since there can be
% different singularities when $t \to 0$ and
%$u \to 0$, which depend on the color basis chosen.

There is already a non-trivial check we can make from just this expression. 
The angular variable $\rnu$ is related to the Mandelstam invariants by Eq.~\eqref{nustu}.
At intermediate states of the calculation 
terms of the form $\ln (\rnu-1)$ appear (see for example Eq.~\eqref{F12eq}). These terms  cannot be expressed simply 
as ratios of Mandelstam invariants, as is required
to cancel the scale dependence of the hard function. The check is that all the $\ln(\rnu -1)$ terms drop out
in the final expression, which involves intricate cancellations among different directions and different color factors.

\subsubsection{Soft RGE}
The RGE for the threshold thrust soft function must be of the general form in Eq.~\eqref{softRGE}.
To see that this equation is satisfied, we first transform to Laplace space via
%---------------------------------------------------------------------------------------
\begin{equation}
\label{Laplace_S}
  \widetilde{S}_{\bI \bJ} (Q,\mu) = 
    \int \rd k \exp
      \left( - \frac{k}{Q e^{\gamma_E}}\right) S_{\bI \bJ} (k,\mu) \,.
\end{equation}
%---------------------------------------------------------------------------------------
For example, for the $qq'\to qq'$ channel, this gives
%---------------------------------------------------------------------------------------
\begin{multline}
\widetilde{S}_{\bI \bJ}(Q, \mu) = 
  \begin{pmatrix}
            \frac{1}{2}C_F C_A      &  0  \\
                      0             &  C_A^2
  \end{pmatrix} 
+\left(\frac{\alpha_s}{4\pi}\right)  c^S_{\bI \bJ}(\frac{t}{u})\\
+\left(\frac{\alpha_s}{4\pi}\right)   
\begin{pmatrix}
        4 C_F\ln\frac{s^2}{ut} - 8 C_F^2 C_A \ln\frac{t}{u} \Theta(u-t)    &~~~ 8 C_A C_F\ln\frac{-u}{s} \\
        8 C_A C_F \ln\frac{-u}{s} &~~~  16 C_A^2 C_F \ln\frac{t}{u} \Theta(t-u) 
\end{pmatrix}\ln\frac{Q}{\mu} \, .
\end{multline}
We have used Eq.~\eqref{nustu} to write the soft function in terms
of $s,t$ and $u$.
We can now check that the soft function for  $qq'\to qq'$  satisfies the RGE, Eq.~\eqref{softRGE}
with
%---------------------------------------------------------------------------------------
\begin{eqnarray}
\label{softqqG}
  \Gamma_{\bI \bJ}^S &=& 
       \left(\frac{\alpha_s}{4\pi}\right) \left(
         \begin{array}{cc}
            -\frac{4}{C_A} ( \ln\frac{tu}{s^2} + 2\pi i ) 
-8 C_F \ln\frac{t}{u}\, \Theta(u-t)
         &  
            \quad8 ( \ln\frac{-u}{s} + i\pi)
         \\
            \frac{4C_F}{C_A}( \ln\frac{-u}{s} + i\pi)
         &  
            8C_F \ln\frac{t}{u}\,\Theta(t-u)
         \end{array}
       \right)   \,.
\end{eqnarray}
%---------------------------------------------------------------------------------------
This matrix is not uniquely defined by the NLO soft function, 
In fact, the soft RGE would also be satisfied for a $\Gamma_{\bI \bJ}^S$ with 
somewhat different off-diagonal terms or imaginary parts. 
However, it is a powerful check on the soft function that it can be written in this way, which
is of the form expected by RG invariance, Eq.~\eqref{GSijgen}. 

For all channels, we find that the soft functions for threshold thrust
satisfy the general RGE, Eq.~\eqref{softRGE}, which takes the particular form
%---------------------------------------------------------------------------------------
\begin{equation} \label{softgenB}
  \Gamma^S_{ {\blue I J}} = 
          \left(
          \gcusp c_Q \ln \left( \frac{Q}{\mu} \right) 
          + \gcusp\rofn
          + \gamma_S 
          \right) \delta_{\bI \bJ} 
          + \gcusp \MIJ \,,
\end{equation}
%---------------------------------------------------------------------------------------
with
%---------------------------------------------------------------------------------------
\begin{align}
\rofn &= c_m\Big[L(t)- \ln\Big(-\min(t,u)\Big)\Big] +\frac{c_Q}{2}[\ln s - L(t)]\\
c_m &= \sum_{\text{initial}} C_{R_i}   \label{cmeq}\\
c_Q &= 
    \sum_{\text{initial}} C_{R_i} 
    - \sum_{\text{final}} C_{R_i} \label{cQeq}\\
\label{r:def}
\gamma_S &= 0 + \cO(\alpha_s^2) \,,
\end{align}
%---------------------------------------------------------------------------------------
with $L(x) = \ln(-x) - i \pi \Theta(x)$, so $L(t) = \ln (-t)$, as in~\cite{Kelley:2010fn}.
For the crossings of $\Gamma^S_{ {\blue I J}}$, direct calculation shows that
 only the $t$ in the $L(t)$ part of $\rofn$ and the $s,t$ and $u$ in $\MIJ$ get crossed. The
$\min(t,u)$ and $s$ factors are the same for all channels.  
To avoid any confusion, $\rofn$ for all crossings
is given explicitly in Tables~\ref{tab:perm1} and~\ref{tab:perm2}.
The $\MIJ$ matrices are the same as for the hard evolution, which 
are given for all channels in~\cite{Kelley:2010fn}.

The appearance of $\min(t,u)$ in $\Gamma^S_{ {\blue IJ }}$
 foreshadows the satisfaction of the RGE consistency check we perform in the next section -- it will
cancel the factor of $\min(t,u)$ in the factorization formula, Eq.~\eqref{ttform}. 
The soft function $\min(t,u)$ factor comes from differences
in the locations of phase space singularities when $t>u$ and when $t<u$, 
which trace back to our convention for how the fermions
are contracted in color space the original operators. 
There can be no factor of $\min(t,u)$ in the hard function, since the hard function 
is analytic and independent of the observable.

To solve the RGE, we diagonalize $\MIJ$. Consistent with~\cite{Kelley:2010fn}, we use the index $\bK$ for
the diagonal basis. The combinations of Wilson lines for which the soft function evolution is diagonal are then $\cW_\bK$,
and we write
\begin{equation}
  \cW_\bK = F_{\bK \bI} \cW_\bI \,,
\end{equation}
and call the eigenvalues $\lambda_\bK$.
Since the same off diagonal terms
appear in the hard and soft evolution, the same basis simultaneously diagonalizes the evolution equation of the hard and soft
functions:
%---------------------------------------------------------------------------------------
\begin{equation} 
H_{\bK \bK'} = (F \cdt H \cdt F^\dagger)_{\bK \bK'} \qquad
S_{\bK \bK'} = [(F^{-1})^\dagger \cdt S \cdt F^{-1}]_{\bK \bK'} \,. 
\end{equation}
%---------------------------------------------------------------------------------------
The eigenvalues, $\lambda_\bK$ and matrices $F_{\bI \bK}$ which diagonalize $\MIJ$ are given for
all channels in~\cite{Kelley:2010fn}. 

For example, for $qq'\to qq'$, 
%---------------------------------------------------------------------------------------
\begin{align} \label{Oqqdiag}
\cW_{\bpm}  &= \lambda_\bpm \cW_\bO+ \frac{C_F}{C_A}\left[\ln \frac{-u}{s} + i \pi\right] \cW_\bT \,,
\end{align}
%---------------------------------------------------------------------------------------
and the eigenvalues are given by
%---------------------------------------------------------------------------------------
\begin{equation} 
\label{evalues}
\lambda_{\bpm} = \frac{C_A}{2}\ln\frac{u}{t}
 - \frac{1}{C_A}\left[\ln \frac{-u}{s} + i \pi\right]
 \bpm \sqrt{\left[\ln \frac{-u}{s} + i \pi\right]\left[\ln \frac{-t}{s} + i \pi\right]+\frac{C_A^2}{4}\ln^2\frac{u}{t} }
\, .
\end{equation}
%---------------------------------------------------------------------------------------
In matrix notation, we have $\cW_{\bpm} = F_{\bpm \bI} \cdt \cW_{\bI}$ where $F$ is
%---------------------------------------------------------------------------------------
\begin{equation} 
\label{Fdef}
F_{\bpm \bI} = 
\begin{pmatrix}
  \lambda_{\blue +} & \quad \frac{C_F}{C_A}\left[\ln \frac{-u}{s} + i \pi\right] \\[2pt]
  \lambda_{\blue -} & \quad \frac{C_F}{C_A}\left[\ln \frac{-u}{s} + i \pi\right]
\end{pmatrix} \,.
\end{equation}
%---------------------------------------------------------------------------------------
The operators and coefficients for the other channels and color structures 
are presented in Appendix~\ref{Appendix_Soft}.

Once the soft function is diagonal, its RGE can be solved in Laplace space, as in Eq.~\eqref{genS}.
In this case, since the soft function has a single scale, we write
\begin{equation}
  \widetilde{S}_{\bK\bK'}(\{ Q \},{\red n_i^\mu},\mu)=
  \widetilde{S}_{\bK\bK'}(\ln\frac{Q}{\mu},\rnu,\mu)
\end{equation}
Then,
%---------------------------------------------------------------------------------------
\begin{align}
&  S_{\bK\bK'}(k,\rnu,\mu) =
     \exp\Big[ -2c_Q  S(\mu_s, \mu)  +2 A_S (\mu_s, \mu)
+A_{\Gamma} (\mu_s, \mu)  
          \left(   
             \lambda_\bK 
            +\lambda_{\bK'}^\star \right)\Big]\\
&\times
\exp\Big[A_{\Gamma} (\mu_s, \mu) \left(2c_m \ln\left|\frac{t}{\min(t,u)}\right| + c_Q\ln\left|\frac{s}{t}\right|\right)\Big]
\widetilde{S}_{\bK\bK'}(\partial_{\eta_s},\rnu,\mu_s) 
\frac{1}{k}\left( \frac{k}{\mu_s}\right)^{\eta_s} 
\frac{e^{- \gamma_E  \eta_{s}}}{\Gamma (\eta_{s})} \,, \nn
\end{align}
%---------------------------------------------------------------------------------------
where $\eta_s = 2C_Q A_\Gamma(\mu_s,\mu)$. For the crossed processes, only $t$ gets crossed, while the $s$ and $\min(t,u)$ factor are
$s$ and $\min(t,u)$ for all channels.
Other examples of this notation can be found in~\cite{Becher:2008cf,Becher:2009qa}.

\subsection{Checking RGE invariance at NLO \label{sec:RG}}
Now that we have computed all the ingredients, we can combine them
together to form the final distribution. Before accounting
for all the kinematic factors, we can
check that the factorization formula, Eq.~\eqref{ttform}, is RG invariant at order $\alpha_s$.
This must hold for each channel and each spin separately.

To check RG invariance,  we first go to Laplace space.
We define the Laplace transform of the cross section as
%---------------------------------------------------------------------------------------
\begin{equation}
\frac{\rd \widetilde\sigma}{\rd Q^2} = \int_0^\infty \rd\tau \, 
\exp \left(-\frac{\tau S}{Q^2 e^{\gamma_E}}\right) \frac{\rd \sigma}{\rd \tau} \,.
\end{equation}
%---------------------------------------------------------------------------------------
Then RG invariance requires, for each channel and spin, that  in the threshold limit,
%---------------------------------------------------------------------------------------
\begin{multline}\label{eq:RGinv}
\frac{\rd}{\rd \ln \mu} \left[
\sum_{\bK, \bK'}s
\widetilde{f}_{\rO} \left( \frac{Q^2}{-\min(t,u)}, \mu \right)
\widetilde{f}_{\rT} \left( \frac{Q^2}{-\min(t,u)}, \mu \right)
\widetilde{j}_\rTH (Q^2,\mu)
\widetilde{j}_\rF (Q^2,\mu)
\right.\\\left. \times \;
H_{\bI \bJ} (\mu)
\widetilde{S}_{\bK'\bK}\left(\frac{Q^2}{\sqrt{s} \mu},\rnu, \mu \right)
\right] = 0\, .
\end{multline}
%---------------------------------------------------------------------------------------
 Plugging in each of the RG equations, this implies
%---------------------------------------------------------------------------------------
\begin{multline}
\gcusp\Big[
2 c_\rO \ln\frac{Q^2}{-\min(t,u)}+ 2 c_\rT \ln\frac{Q^2}{-\min(t,u)} 
-2 c_\rTH \ln\frac{Q^2}{\mu^2} -2 c_\rF \ln\frac{Q^2}{\mu^2}
\\ 
+c_H \ln\frac{-t}{\mu^2} 
+ \lambda_{\bK} + \lambda_{\bK'}^\star
-2 \left( c_Q \ln\frac{Q^2}{\sqrt{s} \mu} 
+ c_m \ln\frac{-t}{-\min(t,u)}
+\frac{c_Q}{2} \ln\frac{s}{-t} \right)
 - \lambda_{\bK} - \lambda_{\bK'}^\star\Big]\\
+2\gamma_{f_\rO} + 2\gamma_{f_\rT} 
-2\gamma_{J_\rTH} - 2\gamma_{J_\rF} 
+2 \gamma_H 
-2 \gamma_S 
= 0 \label{rgcheck} \,.
\end{multline}
%---------------------------------------------------------------------------------------
To check this equation, we need to use that
%---------------------------------------------------------------------------------------
\begin{align}
c_H &= c_\rO + c_\rT + c_\rTH + c_\rF \\
c_Q &= c_\rO + c_\rT-c_\rTH - c_\rF \\
c_m &= c_\rO + c_\rT \,.
\end{align}
%---------------------------------------------------------------------------------------
These relations show that the part of Eq.~\eqref{rgcheck} proportional to $\gcusp$ is satisfied.
The remainder is satisfied if
%---------------------------------------------------------------------------------------
\begin{equation}
  \gamma_S =
\gamma_{f_\rO} + \gamma_{f_\rT} 
-\gamma_{J_\rTH} - \gamma_{J_\rF} 
+\gamma_H  
 \,.
\end{equation}
%---------------------------------------------------------------------------------------
Using the anomalous dimensions we have calculated, this holds at order $\alpha_s$. This equation
can the be used to determine the soft anomalous dimension to $\alpha_s^3$.

\subsection{Threshold Thrust distribution \label{ttfinal}}
The final threshold thrust distribution is achieved by combining the hard, jet
and soft functions and adding appropriate kinematic
factors, as in Eq.~\eqref{ttHJJS}. The result is
%---------------------------------------------------------------------------------------
\begin{multline} 
    \frac{\rd \sigma}{\rd p_T \rd y} = \frac{1 }{8\pi p_T}
\sum_{\text{channels}}
\frac{1}{N_{\text{init}}} 
\int_{\frac{p_T}{\sqrt{S}}e^y}^{1-\frac{p_T}{\sqrt{S}}e^{-y}} \rd v
\int_{\frac{p_T}{\sqrt{S}}\frac{1}{v}e^y}^1 \rd w\frac{v}{w^2}
        \left[x_\rO f_{\rO/N_1} (x_\rO,\mu)\right]
        \left[x_\rT f_{\rT/N_2} (x_\rT,\mu)\right]\\
\times
\sum_{\bK, \bK'}
  \frac{\alpha_s(\mu_h)^2}{\alpha_s(\mu)^2} 
H_{\bK \bK'}(v,\mu_h) 
\\ \times
        \exp \left[ 
             2 c_H S (\mu_h, \mu)
            +2 A_H (\mu_h, \mu) 
            -2 c_Q S(\mu_s, \mu)
            +2 A_S(\mu_s, \mu)
\right]\\
~~~~~~~~~~~~~~~~~~~~~\times \exp \left[
            -4 c_{\rTH} S (\mu_j, \mu) 
            +2 A_{J_\rTH} (\mu_j, \mu)
            -4 c_{\rF} S (\mu_j, \mu) 
            +2 A_{J_\rF} (\mu_j, \mu)
\right]\\
~~~~~~~~~~~~~~~~~~~~~~~~    \times \exp\left[  
      A_{\Gamma} (\mu_s, \mu_h) 
             \left( \lambda_\bK + \lambda_{\bK'}^\star\right) 
      -A_{\Gamma} (\mu_h, \mu) c_H \ln  \left| \frac{t}{\mu_h^2} \right| 
\right]\\
\times \exp \left[
     A_{\Gamma} (\mu_s, \mu) \left(
     2c_m\ln\left|\frac{t}{p_T^2}\min(v,\vb)\right|
     + c_Q  \ln\left| \frac{\mu_j^4/\mu_s^2}{ t}\right|\right)
             \right]
\\\times
    \widetilde{j}_{\rTH}(\partial_\reta,\mu_j)
    \widetilde{j}_{\rF}(\partial_\reta,\mu_j)
\widetilde{S}_{\bK\bK'}(\frac{\mu_j^2v\vb}{\mu_s p_T}+\partial_\reta,\frac{\vb}{v},\mu_s)
\left[\frac{1}{s_4}\left(\frac{s_4}{\mu_j^2}\right)^\reta\right]_\star^{[s_4,\mu_j^2]}
    \frac{e^{- \gamma_E \reta}}{\Gamma (\reta)}  \,.
\end{multline}
%---------------------------------------------------------------------------------------
Here, $c_\rTH$ and $c_\rF$ are the fundamental Casimirs ($C_F$ or $C_A$) for the $\rTH$ or $\rF$ jet
and
\begin{equation}
  \reta =  \eta_\rTH + \eta_\rF+\eta_s = 2(c_{\rTH}+c_\rF) A_\Gamma(\mu_j,\mu)+2c_Q A_\Gamma(\mu_s,\mu)
\end{equation}
We have also used
\begin{equation}
  |\min(t,u)| = p_T^2 \frac{1}{\min(v,\vb)}
\end{equation}
and  $\rnu = \frac{t}{u} = \frac{\vb}{v}$ to write things directly in terms of $v$. 
For the different channels, one must use the appropriately crossed hard, jet and soft functions and PDFs,
and additionally cross the explicit factors of $t$. 
%The hard and soft functions
%also depend on 
%whether we set $w=1$ in the hard functions (so that $s = \frac{p_T^2}{v(1-v)}$, $t=-\frac{p_T^2}{v}$, and $u=-t-s$) is a convention.
%This conventional will produce different power corrections from the point of view of SCET, which are removed in matching to the
%fixed order distribution, as in~\cite{Becher:2009th}.

\section{Dynamical Threshold Enhancement \label{sec:dynamic}}
The threshold thrust variable we have calculated in this paper allows us to calculate the $p_T$ and rapidity distribution of
dijet events jets near the machine threshold. In this threshold region, the final state consists of two almost massless jets and nothing
else. Such a threshold is physically irrelevant because the final state in any realistic collision has beam remnants
which contribute to the hemisphere masses, preventing them from ever being close to massless.
However, since the beam remnants have no transverse momentum, the $p_T$ distribution of realistic jets
should have singular contributions very similar to the hemisphere jets in threshold thrust.
Thus, it should be possible to calculate the physical jet $p_T$ spectrum using SCET and to compare directly to
data.

The reason the SCET factorization theorem is applicable only in the threshold region is because in that region
the momenta of the incoming partons approach the momenta of the incoming hadrons. In this limit a hadron can be thought of
as a parton with the other, spectator, partons in the hadron providing only power suppressed contributions.
Thus, the singular logarithmic terms
are guaranteed to be the dominant contribution to the cross section at threshold.
For the large logs, and their resummation, to be important away from threshold, their relative contribution must
be somehow enhanced from what one might naively expect. This is called dynamical threshold enhancement.

Dynamical threshold enhancement has been shown to hold for other processes
such as Drell-Yan and direct photon production. In Drell-Yan, the observable is the mass of the final state leptons, $m^2_{ll} = (p^\mu_{l^+} + p^\mu_{l^-})^2$ and the threshold region is when $ m^2_{ll} \to S$. In Direct photon, the observable is the mass of 
everything-but-the-photon, $M_X^2 = S+T+U$ and the threshold region is when $M_X^2 \to 0$.
Dynamical threshold enhancement in direct photon can be understood intuitively. $M_X$ includes
contributions from the mass of the final state jet. Away from threshold,
the factorization theorem does not guarantee that the jet mass would be small compared to its energy. However, we know
from observation that the typical jet mass in a realistic event is in fact small. Therefore, the large logarithms at threshold
related to the jet mass are also relevant in the physical regime. 

For threshold thrust, the observable $\tau$ is very similar to the observable for direct photon,
since $\tau =( S + T +U)/S$. We expect large logarithms of the jet mass to be similarly important even away from threshold,
since typical jet masses are small compared to their energy in both cases. However, there is an important difference. For direct photon,
we can write the final state momentum as $p_{\text{final}}^\mu = p_\gamma^\mu + p_X^\mu$. Since $p_X^\mu$ is the momentum of everything
but the photon, it includes both beam remnants. Thus, the direction of $p_X^\mu$ is the same at the parton and hadron level,
even though its energy may be vastly different when $s \ll S$. 
That the direction is the same is critical for SCET, because the direction is a label for the operator, 
and appears in the final distribution. In contrast, the energy of the final state jet cancels out
at an intermediate stage of the calculation, and only reappears through the matching scales. 

For dijet events, we can write the final state as 
$p_{\text{final}}^\mu =p_\rR^\mu + p_\rL^\mu$. 
Since one beam remnant goes into one hemisphere and the other beam remnant goes into the other hemisphere, 
the direction of the final states at the parton and hadron levels can be vastly different. In other words,
the hemisphere axis is sensitive to the boost of the event, while the direction of the jet in direct photon
is not. If the direction label on the jet changes, the matching coefficients should change too. So, there
is no reason to expect the threshold thrust calculation to be relevant in the physical regime. 

Suppose we try to modify the definition of the observable so that its direction is stable away from threshold. For example,
we could incorporate a jet definition, say a cone of size $R$. Then instead of threshold thrust,
which is the sum of hemisphere masses-squared, we look at the sum of the squares of the masses
of the two hardest jets of size $R$ in the event, $\tau_{J^2} = (p^\mu_\rR)^2 + (p^\mu_\rL)^2$.
Since these jets would not include the beam, they should be stable
between the parton and hadron levels. This modified observable has (at least) two problems. First, it is not global, so there
may be relevant non-global logs which we are not resumming, even at threshold. Second, the vanishing of the observable does
not guarantee the dijet kinematics relevant to the factorization theorem. While the observable vanishing does imply that
both jet masses must vanish, it does not enforce that the remaining radiation should be soft. For example, there could be
other hard jets in the event. Thus, while the observable is good away from threshold, it is not useful {\it at} threshold.

In summary, there appear to be two general principles which an observable should satisfy for dynamical
threshold enhancement to work. We would like to have a single definition of the observable which 
can be applied both at threshold and away from threshold so that
\begin{enumerate}
\item The direction of the final state jets is the same at the parton and hadron levels. \label{itemone}
\item The vanishing of the observable implies the kinematics associated with the factorization theorem. \label{itemtwo}
\end{enumerate}
In the case of hadronic dijet event shapes, the second point means that only $2\to 2$ processes should contribute when
the observable is exactly at threshold.
For Drell-Yan, both of these criteria hold: there are no final state jets, and when $m^2_{ll} \to S$, the remaining
radiation must be soft. For direct photon, these criteria hold: the jet defined as the everything-but-the-photon is always
back-to-back with the photon, and $M_X\to 0$ forces the jet to be massless and the remaining radiation to be soft.
Threshold thrust satisfies the second condition but  fails the first because the beam remnants change the jet direction.

In addition, to avoid complications involving non-global logs
\begin{enumerate}
\setcounter{enumi}{2}
\item The observable should be global.\label{itemthree}
\end{enumerate}
This means all hadrons in the event should contribute to the observable~\cite{Dasgupta:2001sh,Dasgupta:2002dc}.
For direct photon and Drell-Yan, the observables are the photon 4-momentum  or lepton pair 4-momentum respectively,
which seem clearly not to be global since they do not involve the hadrons at all.
However, in direct photon, what is actually calculated in SCET is
the mass of everything-but-the-photon, $M_X$, which is global. In Drell-Yan, one can also rephrase the calculation
as being of the mass of everything but the lepton pair, $M_X^2 = S + T + U - m_{ll}^2$, which is global as well. For threshold
thrust, $M_X^2 = S+T+U = P_\rR^2 +P_\rL^2$, which is also global. This is not to say that observables which are non-global
must have non-global logarithms, or that non-global logarithms cannot be resummed. Moreover, there are additional subtleties
within the class of global observables, such as whether they are directly or continuously global~\cite{Dasgupta:2002dc}.
So it makes sense to start in SCET by studying
event shapes which are as global as possible, so that all the large logarithms will be threshold logs.

Now let us consider some modifications of threshold thrust. First, take the example mentioned above, the sum of the
squares of the masses of the two hardest jets of size $R$, $\tau_{J^2}= (p^\mu_\rL)^2 + (p^\mu_\rR)^2$.
It will satisfy condition~\ref{itemone} but not condition~\ref{itemtwo}.
Even though the jets are forced to be massless when $\tau_{J^2}=0$, the radiation outside the jets is not forced
to vanish by $\tau_{J^2} = 0$ alone. 
 The variable $\tau_{+} =\frac{1}{S}(p_\rR^\mu + p_\rL^\mu)^2$, discussed in~\cite{Laenen:1998qw} and~\cite{Bauer:2010vu}
has the the opposite problem. In this case, $\tau_{+} = 1$ does force the radiation outside the jets to vanish,
however, the jets are not forced to be massless. They can have masses as large as $m_J /E_J \lesssim R$. Although this
observable may be interesting, resumming logs of the jet masses cannot be done without also resumming logs of $R$.
If $R$ is as large as a hemisphere, then $\tau_{+} =1$ exactly, so there is nothing to calculate.
 A related observable,
suggested by~\cite{Laenen:1998qw} and studied in~\cite{deFlorian:2007fv} is  $\tau_{-} =\frac{1}{S}(2 p_\rR\cdt p_\rL) = \tau_{+}-\tau_{J^2}$. 
When this observable is at threshold, $\tau_{-}=1$, the final state is
forced to have two massless jets and no out of jet radiation. It satisfies both criteria~\ref{itemone} and~\ref{itemtwo},
 and therefore we expect it to 
undergo dynamical threshold enhancement. However, it is not global: because the jets have size $R$, only the radiation within
the jets contributes to the value of $p_\rR\cdt p_\rL$. Unlike direct photon or Drell-Yan, the mass of everything not in the
jets is also non-global beacuse it excludes the hadrons in the jets.  Thus, when attempting NNLL resummation,
there may be non-global logs in $\rd \sigma/\rd \tau_{-}$ of the same order as the logs which are resummed.

Finally, consider the following modification of threshold thrust. Instead of taking two hemispheres, one finds the
hardest jet of size $R$ (using whatever algorithm one wants). Then the radiation in the event is divided into
radiation in the jet, which is summed into a momentum 4-vector $p_{{\red \text{in}}}^\mu$ and out of the jet,
which is summed into $p_{{\red \text{out}}}^\mu$. Then one computes the asymmetric thrust
\begin{equation}
\tau_A =\frac{1}{S}\left[ ( p_{{\red \text{in}}}^\mu)^2 + (p_{{\red \text{out}}}^\mu)^2\right] \,.
\end{equation}
This will allow us to resum logarithms of the $p_T$ of the jet, which can be compared to data and should undergo threshold
enhancement. The exact $R$-dependence can be included through careful matching to a fixed order calculation.

For $R$ the size of a hemisphere, this observable is just threshold thrust. For $R$ small enough
so that neither beam is included in the jet, this observable will satisfy condition~\ref{itemone}. In fact, if
$R$ is calculated as a distance in $\eta - \phi$ space, the beams will never be included in a jet since they are at $\eta=\pm\infty$.
Condition~\ref{itemtwo} is satisfied as well. When $M_X\to 0$, both the jet mass $\sqrt{(p_{{\red \text{in}}}^\mu)^2}$ and
the mass-of-everything-else $\sqrt{(p_{{\red \text{out}}}^\mu)^2}$ are forced to vanish. The initial state radiation must
also be soft, for the same reason as for threshold thrust.
 Thus, at threshold, only $2\to 2$
scattering contributes, as desired. Finally, the observable is global, so there are no dangerous non-global logs
to worry about.

%#######################################################################################3
%#######################################################################################3
%#######################################################################################3
%#######################################################################################3

\section{Conclusions \label{sec:conc}}
In this paper, we have presented the first computation of a hadronic event shape in pure QCD events
to next-to-next-to-leading logarithmic accuracy. This calculation of this observable, which 
we call threshold thrust,  involves understanding
a number of issues which were not present in previous work. The hadronic event shapes we have been concerned
with here are dominated by dijet configurations. In the singular limit, at threshold, only $2\to2$ scattering contributes,
but there are many channels and many color structures in each channel. The evolution in color space is complicated,
but has a number of simplifying features. In particular, there is a color basis in which the evolution is diagonal. 
The evolution of the hard functions was studied previously in~\cite{Kelley:2010fn}, and constraints on the RG
evolution for the soft function were derived. Here, we have computed the soft function for threshold thrust
explicitly to 1-loop. This both confirms the general expectations from~\cite{Kelley:2010fn}, and provides
the NLO finite parts of the soft function necessary for NNLL resummation.

Although threshold thrust is infrared safe and observable in principle, the large logarithmic contributions to
it which the SCET calculation provides are unlikely to have a significant effect in the physically relevant
kinematic region. The threshold where the calculation is valid has $x\to 1$ for both hadrons, so that the hadron
momenta and the hard scattering parton momenta approach each other. It is not unusual for the threshold logarithms
to be important away from threshold. In fact, for related processes, such as Drell-Yan and direct photon production,
this is known to happen due to dynamical threshold enhancement. We isolate two features of observables
which are likely to be important for dynamical threshold enhancement to occur: 1) The jets in the observable
should have the same directions at the hadron and parton levels and 2) The observable should force the kinematics
of the factorization theorem. In addition, to avoid non-global logs, 3) The observable should be sensitive
to radiation everywhere. We defined an alternative to threshold thrust, asymmetric thrust, which satisfies these
criteria. While threshold thrust is the sum the masses-squared of hemisphere jets, asymmetric thrust sums the
masses-squared of a jet and everything-but-the-jet. The critical feature of this modification is that both beams
end up in one side, so their contributions to the threshold observable largely cancel. It will be interesting to
study the phenomenology of asymmetric thrust, which, after matching to the exact NLO distribution,
should allow us to compute the jet $p_T$ spectrum at NNLL+NLO accuracy. 

Many of the results and lessons from this paper will help us understand more general hadronic event shapes.
While the $p_T$ of a jet in dijet events is an important observable, there are many other observables which one could
compute using the same $2\to2$ hard kinematics which would provide complimentary tests of QCD.
Some of the global observables discussed in~\cite{Banfi:2010xy}, or 2-jettiness which has been proposed 
using SCET~\cite{Stewart:2010tn},
may be calculable at NNLL using the results of this paper and of~\cite{Kelley:2010fn} without too much more work.
There are also many interesting non-global observables which can be studied and measured using the same dijet sample. 
For example, it would be nice to calculate the mass
of a jet in dijet events. Studies in $e^+e^-$ collisions of thrust~\cite{Becher:2008cf,Abbate:2010xh}, which is the sum
of jet masses, and the related heavy jet mass~\cite{Chien:2010kc}, 
have shown clearly that resummation at NNLL and beyond is critical to getting some shapes correct. 
Because of the beam remnants, a jet mass distribution at hadron colliders requires a less inclusive jet definition than at $e^+e^-$ colliders.
 Much progress has already been made studying
exclusive jet definitions in SCET~\cite{Cheung:2009sg,Ellis:2010rw,Bauer:2010vu}, and phenomenological applications at hadron
colliders are surely just around the corner. However, to go to NNLL requires an understanding of not only non-global logs in SCET,
but also of super-leading logs~\cite{Forshaw:2006fk}, which arise as violations of color coherence in processes with at least four
hard colored particles, such as the dijet configurations we have been studying. It will be interesting to see if SCET can provide
a consistent framework for resummation and precision calculations of general non-global observables at hadron colliders.

\section{Acknowledgements}
We would like to thank Thomas Becher and Gavin Salam for helpful discussions. Our research is
supported in part by the Department of Energy under Grant DE-SC003916.

\appendix

\section{Threshold thrust soft function\label{Appendix_Soft}}
The calculation of the soft function involves taking matrix elements of Wilson
lines. To order $\alpha_s$ in dimensional regularization only the real emission
diagrams contribute. These can be drawn as cut graphs, as in Figure~\ref{fig:soft}.
There are contributions where the gluon comes out of the ${\red n_a}$ leg and is absorbed into
the ${\red n_b}$ leg, where  ${\red n_a}$ and ${\red n_b}$ are any of 
${\red n_1}, {\red n_2},{\red n_\rTH}$ or ${\red n_\rF}$.
Since the graphs with ${\red n_a}={\red n_b}$ vanish, there are 6 possibilities.
The calculation can be split into two parts: calculation of the
integrals, which we call $I_S$, and calculation of the color factors, which we call $D_{\bI \bJ}$. The result can then be written as
%---------------------------------------------------------------------------------------
\begin{equation} 
S_{\bI \bJ}(k_\rR,k_\rL) 
= D_{\bI \bJ}^{\text{tree}}\delta(k_\rTH)\delta(k_\rF)+
  \sum_{\ra, \rb}
  I_S( {\red n_a}, {\red n_b},k_\rTH,k_\rF ) 
  D_{\bI \bJ}(\chi(\ra),\chi(\rb))+\cO(\alpha_s^2) \,.
\end{equation}
%---------------------------------------------------------------------------------------
The notation $\chi(\ra)$ and $\chi(\rb)$ indicates to permute the $\ra$ and $\rb$ indices in the color factor, $\DIJ$,
with respect to the integral, $I_S$, for the appropriate crossing.
The crossings and permutations are listed in Tables~\ref{tab:perm1} and \ref{tab:perm2}. We work in $d=4-2\e$ dimensions with $\overline{\text{MS}}$ subtraction throughout.

\subsection{Integrals} \label{integrals}
The master integral for the soft function is
%---------------------------------------------------------------------------------------
\begin{align}
& I_S ( {\red n_a}, {\red n_b}, k_\rTH, k_\rF) =  
g_s^2
 \left( \frac{ \mu^{2} e^{\gamma_E} }{4\pi} \right)^\e 
    \int \frac{\rd^d q}{(2\pi)^{d-1} } \Theta (q^0) \delta (q^2) 
    \frac{{\red n_a} \cdt {\red n_b}}{( {\red n_a} \cdt q) ( {\red n_b} \cdt q)}
    \nonumber \\
& \quad \times  
    \Bigl\{
        \Theta ( {\red n_\rTH} \cdt q - {\red n_\rF} \cdt q)
        \delta (k_\rF -{\red n_\rF} \cdt q) 
        \delta (k_\rTH ) 
        + 
        \Theta ({\red n_\rF} \cdt q - {\red n_\rTH} \cdt q) 
        \delta (k_\rTH - {\red n_\rTH} \cdt q) 
        \delta (k_\rF ) 
    \Bigr \} \,,
\end{align}
%---------------------------------------------------------------------------------------
where ${\red n_a}$ and ${\red n_b}$ are the directions of the two
Wilson lines where the gluon attaches, which can be any of 
${\red n_1}$, ${\red n_2}$, ${\red n_\rTH}$ or ${\red n_\rF}$, as in Figure~\ref{fig:soft}. Thus, there are six different integrals to be done.
As discussed in the text, the integrals can only depend on one Lorentz invariant ratio
%---------------------------------------------------------------------------------------
\begin{equation}
 \rnu = \frac{{\red n_1} \cdt {\red n_\rTH}}{{\red n_1} \cdt {\red n_\rF}} 
=\frac{{\red n_2} \cdt {\red n_\rF}}{{\red n_2} \cdt {\red n_\rTH}}  \,,
\end{equation}
%---------------------------------------------------------------------------------------
which greatly simplifies the calculations.

At order $\alpha_s$ the soft gluon can either go into the hemisphere containing Wilson line $\rTH$,
or Wilson line $\rF$.
Thus, at order $\alpha_s$ the master integral can be written as
%---------------------------------------------------------------------------------------
\begin{align}
I_S ( {\red n_a}, {\red n_b}, k_\rTH, k_\rF) &= \left(\frac{\alpha_s}{4\pi}\right)
\left[
 \frac{\mu^{2\varepsilon}}{k_\rF^{1 + 2 \varepsilon}}  \delta (k_\rTH) 
L_{\red ab}(\rnu)+
 \frac{\mu^{2\varepsilon}}{k_\rTH^{1 + 2 \varepsilon}} \delta (k_\rF)
R_{\red ab}(\rnu) 
\right] \,,
\end{align}
%---------------------------------------------------------------------------------------
with
%---------------------------------------------------------------------------------------
\begin{align} \label{R_def}
L_{\red a b}(\rnu)
 &=  
    \frac{16\pi^2}{\ell^{-1-2\e}}
    \left( \frac{ e^{\gamma_E} }{4\pi} \right)^\e 
    \int \frac{\rd^d q}{(2\pi)^{d-1} } \Theta (q_0) \delta (q^2) 
    \frac{{\red n_a} \cdt {\red n_b}}{( {\red n_a} \cdt q) ( {\red n_b} \cdt q)}
        \Theta ( {\red n_\rTH} \cdt q - {\red n_\rF} \cdt q)
        \delta (\ell -{\red n_\rF} \cdt q) \,,
\end{align}
%---------------------------------------------------------------------------------------
and 
%---------------------------------------------------------------------------------------
\begin{align}
  R_{\red 1 2}  (\rnu) &= L_{\red 1 2}   (\frac{1}{\rnu}),\quad R_{\red \rTH \rF}(\rnu) = L_{\red \rTH \rF}(\frac{1}{\rnu})\\
  R_{\red 1 \rTH}(\rnu)  &= L_{\red 1 \rF}(\frac{1}{\rnu}),\quad R_{\red 1 \rF}  (\rnu) = L_{\red 1 \rTH}  (\frac{1}{\rnu})\\
  R_{\red 2 \rTH}(\rnu)  &= L_{\red 2 \rF}(\frac{1}{\rnu}),\quad R_{\red 2 \rF}  (\rnu) = L_{\red 2 \rTH}  (\frac{1}{\rnu})
\,.
\end{align}
%---------------------------------------------------------------------------------------
These integrals do not depend on $\ell$, by dimensional analysis, but some expressions tend to appear simpler
before rescaling $\ell$ away. We now proceed to 
 calculate the 6 dimensionless integrals $L_{\ra \rb}(\rnu)$.

The integrals are easiest to do in a basis where the $\delta$ and $\Theta$ functions 
are easy to integrate, which means choosing lightlike coordinates along the $\rTH$ and $\rF$ axis. So we write
%---------------------------------------------------------------------------------------
\begin{equation}
 q^{\mu} = \frac{1}{2} q_- {\red n_\rF^{\mu}} + \frac{1}{2} q_+
 {\red n_\rTH^{\mu}} + q_{\perp}^{\mu} \,.
\end{equation}
%---------------------------------------------------------------------------------------
In this basis the measure, including the $\delta$ and $\Theta$ functions, becomes
%---------------------------------------------------------------------------------------
\begin{align}
        \int \rd^d q \Theta (q_0) \delta (q^2) 
        \Theta ( {\red n_\rTH} \cdt q - {\red n_\rF} \cdt q)
        \delta (k_\rF -{\red n_\rF} \cdt q) \{ \cdots \! \} 
%    \qquad \\       
=
        \frac{\ell^{1-2\e}}{4 }
        \Omega_{d-3} 
        \int_{1}^{\infty} \frac{\rd x}{x^\e}
        \int_{0}^{\pi} \frac{\rd\theta}{\sin^{2\e} \theta} \{ \cdots \!\}
\,,
\end{align}
where $\Omega_d = 2\pi^{d/2}/\Gamma(\frac{d}{2})$ and
\begin{align}
x = \frac{ {\red n_\rTH} \cdt q} \ell \, ,
\end{align}
%---------------------------------------------------------------------------------------
and $\theta$ is the angle between $q_\perp$ and $\vec{\red n}_{\rO \perp}$. 
Then,
%---------------------------------------------------------------------------------------
\begin{equation}
    L_{\red a b}(\rnu) = \frac{2}{\pi}
    \left(  e^{\gamma_E} \pi \right)^\e 
    \Omega_{d-3} 
    \frac{\ell^2}{4}
    \int_{1}^{\infty} \frac{\rd x}{x^\e}
    \int_{0}^{\pi} \frac{d\theta}{\sin^{2\e} \theta} 
    \frac{{\red n_a} \cdt {\red n_b}}{( {\red n_a} \cdt q) ( {\red n_b} \cdt q)}
\,.
\end{equation}
%---------------------------------------------------------------------------------------
The inner products that appear in the denominator are, in this basis,
%---------------------------------------------------------------------------------------
\begin{align}
    {\red n_1} \cdt q &= 
    \frac{\ell}{1+\rnu} 
    \Bigl\{ 
            x + \rnu- 2 \sqrt{\rnu x} \cos \theta 
    \Bigr\} \\
    {\red n_2} \cdt q &= 
    \frac{\ell}{1+\rnu} 
    \Bigl\{ 
            x \rnu +1 + 2\sqrt{\rnu x} \cos \theta 
    \Bigr\} \\
    {\red n_\rTH} \cdt q &= q_- \equiv \ell x\\
    {\red n_\rF} \cdt q &= q_+ = \ell  \,.
\end{align}
%---------------------------------------------------------------------------------------
Since $\rn_\rO = \overline{\rn}_\rT$, 
the angle between $q_\perp$ and $\vec{\rn}_{\rT \perp}$ is $\pi - \theta$ 
thus changing the sign of the $\cos \theta$ term in the expression for $\rn_\rT \cdt q$.
The terms $1+\rnu$ arise in the center of mass frame where $\rn_\rTH + \rn_\rF = (2,\vec{0})$ since
%---------------------------------------------------------------------------------------
\begin{equation}
1+ \rnu 
    = 1 + \frac{(\rn_\rO \cdt \rn_\rTH)}{(\rn_\rO \cdt \rn_\rF)} 
    = \frac{2}{(\rn_\rO \cdt \rn_\rF)}
    = \frac{2}{(\rn_\rT \cdt \rn_\rTH)} \,.
\end{equation}
%---------------------------------------------------------------------------------------

There are 6 ways to choose two of the four directions for ${\red n_a}$ and
${\red n_b}$, with $\rn_\ra \neq \rn_\rb$. These integrals are tricky, but can be
done in a straightforward manner. Doing the $\theta$ integral first leads to a
form involving a hypergeometric function. The $x$ integral then may have singularities at
$x = 0, x = 1, x = \rnu $ and $x = \infty$. Also, when $\rnu = 1$, the ${\red n^\mu_\rO}$ and  ${\red n^\mu_\rT}$ 
directions are on the hemisphere boundary.
Thus, special care is required around $\rnu=1$ and it turns out that the regions
with $\rnu>1$ and $\rnu <1$ have different functional dependence on $\rnu$.
All of the singularities are regulated by $\e$, but special care must be
taken to properly treat the various branch cuts. It is helpful also to use some tricks similar to those
described in~\cite{Ellis:2010rw}.

\subsubsection*{case ${\red n_a} = {\red n_\rTH}$, ${\red n_b} = {\red n_\rF}$  }
This case is the same as for the $e^+e^-$ hemisphere mass distribution~\cite{Schwartz:2007ib,Fleming:2007xt}. 
%---------------------------------------------------------------------------------------
\begin{equation}
    L_{\rTH \rF}(\rnu ) = 
    \frac{\left(  e^{\gamma_E} \pi \right)^\e }{\pi} 
    \Omega_{d-2} 
    \int_{1}^{\infty} \frac{\rd x}{x^{1+\e}}
\,.
\end{equation}
%---------------------------------------------------------------------------------------
Expanding to order $\e$ gives
%---------------------------------------------------------------------------------------
\begin{align}
L_{\red \rTH \rF}(\rnu)  =     \frac{2}{\e} - \e \frac{\pi^2}{6}  
\,.
\end{align}
%---------------------------------------------------------------------------------------

\subsubsection*{case: ${\red n_a} = {\red n_1}$, ${\red n_b} = {\red n_2}$ }
For this case,
%---------------------------------------------------------------------------------------
\begin{align}
L_{\rO \rT}(\rnu) 
&= 
    \frac{\left(  e^{\gamma_E} \pi \right)^\e }{\pi} 
    \Omega_{d-3} 
    (1+\rnu)^2
    \nn \\
&    
    \times
    \int_{1}^{\infty}\!\! \rd x
    \int_{0}^{\pi}\! \frac{\rd \theta}{\sin^{2\e} \theta} 
    \frac{x^{-\e}}{( x + \rnu- 2 \sqrt{\rnu x} \cos \theta ) ( x \rnu +1 + 2\sqrt{\rnu x} \cos \theta)}
\,.
\end{align}
%---------------------------------------------------------------------------------------
This integral is poorly behaved at $\rnu = 1$, where ${\red n_\rO^\mu}$ and  ${\red n_\rT^\mu}$ 
are on the hemisphere boundary. However, the integral is symmetric in $\rnu \to \frac{1}{\rnu}$,
so we can write the result as
%---------------------------------------------------------------------------------------
\begin{align}
L_{\rO \rT}(\rnu) 
= 
        F_{\rO \rT}( \rnu      ) \Theta ( \rnu - 1 )  
        +
        F_{\rO \rT}\left( \frac{1}{\rnu}\right) \Theta (1 - \rnu)\,,
\end{align}
%---------------------------------------------------------------------------------------
and just perform the integral assuming $\rnu >1$. The result is
%---------------------------------------------------------------------------------------
\begin{multline} \label{F12eq}
F_{\rO \rT}( \rnu ) = 
    -\frac{2}{\varepsilon} 
        + 4 \ln \left( 1 + \rnu \right)
        +2\e 
        \left( 
            -3 \LI\left(\frac{1}{\rnu}\right)
            -2 \LI\left(1-\rnu\right) 
        \right. \\
        \left.
                -2\ln ^2\left(\rnu+1\right)
                - \ln \left(\rnu -1\right)\ln\rnu
                +2\ln \left(\rnu+1\right)\ln\rnu
                +2\ln ^2(\rnu )
            -\frac{5 \pi ^2}{12} 
        \right) \,.
\end{multline}
Note that $L_{\rO \rT}(\rnu)$ is real for all $\rnu$.
%---------------------------------------------------------------------------------------
\subsubsection*{case: ${\red n_a} = {\red n_1},{\red n_2}$, ${\red n_b} = {\red n_\rTH}, {\red n_\rF}$  }
%---------------------------------------------------------------------------------------
This is the most complicated case and we will start with $\rn_\ra = \rn_\rO$. 
The integrals are most easily evaluated using the same light cone basis as in previous integral.  Then,
%---------------------------------------------------------------------------------------
\begin{align}
L_{\rO \rTH}(\rnu) 
&= 
    \frac{\left(  e^{\gamma_E} \pi \right)^\e}{\pi}
    \Omega_{d-3} 
    \int_{1}^{\infty}\!\! \rd x\frac{\rnu}{x}\,
    \int_{0}^{\pi}\! \frac{\rd \theta}{\sin^{2\e} \theta} 
    \frac{x^{-\e}}{( x + \rnu- 2 \sqrt{ \rnu x} \cos \theta ) } 
\\
L_{\rO \rF}(\rnu) 
&= 
    \frac{\left(  e^{\gamma_E} \pi \right)^\e}{\pi} 
    \Omega_{d-3} 
    \int_{1}^{\infty}\!\! \rd x
    \int_{0}^{\pi}\! \frac{\rd \theta}{\sin^{2\e} \theta} 
    \frac{x^{-\e}}{( x + \rnu- 2 \sqrt{\rnu x} \cos \theta ) } \,.
\end{align}
%---------------------------------------------------------------------------------------
The results are written in terms of four functions in order to account for the separate cases 
$\rnu > 1$ or $\rnu < 1$. We find
%---------------------------------------------------------------------------------------
\begin{align}
L_{\rO \rTH}(\rnu) &= 
        F_{\rO \rTH}\left( \rnu           \right) \Theta ( \rnu - 1 )  
        +
        G_{\rO \rTH}\left( \rnu \right) \Theta (1 - \rnu) \\
L_{\red 1\rF}(\rnu) &= 
        F_{\rO \rF}\left( \rnu           \right) \Theta ( \rnu - 1 )  
        +
        G_{\rO \rF}\left( \rnu \right) \Theta (1 - \rnu)\,,
\end{align}
%---------------------------------------------------------------------------------------
where
%---------------------------------------------------------------------------------------
\begin{eqnarray}
F_{\rO \rTH}( \rnu ) &=& 
    -\frac{2}{\varepsilon} 
        + 2 \ln \left( \rnu \right)
        + 2 \ln \left( \rnu - 1 \right)
\nn \\
&&
        +2\e 
        \left( 
            -3 \LI\left(\frac{1}{\rnu}\right)
            - \ln^2( \rnu -1 ) 
            +2 \ln  ( \rnu -1 ) \ln( \rnu ) 
            -2 \ln^2 ( \rnu ) 
            + \frac{\pi^2}{12}
        \right) 
\nn \\
G_{\rO \rTH}( \rnu ) 
&=& 
        - 2 \ln \left( 1 - \rnu  \right)
        +2\e 
        \left(\phantom{\frac{1}{\rnu}}\!\!\!\! 
            \ln^2(1 -  \rnu ) 
           + 2\ln  (1 -  \rnu )  \ln  ( \rnu  ) 
          \right) 
\nn \\
F_{\rO \rF}( \rnu ) 
&=& 
        2 \ln \left( \frac{\rnu - 1}{\rnu}  \right)
        +2\e 
        \left( 
            -\LI\left(\frac{1}{\rnu}\right)
            -
            \ln^2 \left( \frac{\rnu - 1}{\rnu}  \right)
        \right) 
\nn \\
G_{\rO \rF}( \rnu ) 
&=& 
        \frac{2}{\e} 
        - 2 \ln \left( 1 - \rnu  \right)
        +2\e 
        \left( 
            \LI( \rnu )
            +\ln^2 (1 -  \rnu ) 
            -\frac{\pi^2}{12} 
         \right) \,.
\end{eqnarray}
%---------------------------------------------------------------------------------------
The case $\rn_\ra = \rn_\rT$ can be expressed in terms of the same integral as the 
case $\rn_\ra = \rn_\rO$.
%---------------------------------------------------------------------------------------
\begin{align}
L_{\red 2\rF}(\rnu) = 
L_{\red 1\rF}(\frac{1}{\rnu})&=
        G_{\rO\rF}\left( \frac{1}{\rnu} \right) \Theta ( \rnu - 1 )  
        +
        F_{\rO \rF}\left( \frac{1}{\rnu} \right) \Theta (1 - \rnu)\\
L_{\red 2\rTH}(\rnu) =
L_{\red 1\rTH}(\frac{1}{\rnu})&=
       G_{\rO \rTH}\left( \frac{1}{\rnu} \right) \Theta ( \rnu - 1 )  
       +
       F_{\rO\rTH}\left( \frac{1}{\rnu} \right) \Theta (1 - \rnu) \,.
\end{align}
%---------------------------------------------------------------------------------------

\subsection{color factors \label{Appendix_Color}}
The color factors that accompany each integral depend on the channel and the crossing.
They are particular to the basis of operators used and the associated color structures 
in the soft functions. We only need to calculate the color factors $\DIJ(\ra, \rb)$ for the
three representative channels ($qq'\to qq', gg \to q \bar{q}$ and $gg \to gg$), with the crossed processes determined by permutations
of the indices, as described in the text and in Tables~\ref{tab:perm1} and~\ref{tab:perm2}.

%---------------------------------------------------------------------------------------
%------------------------ qQ -> qQ -----------------------------------------------------
%---------------------------------------------------------------------------------------
\subsubsection*{case $ q q^\prime \to q q^{\prime} $ } 
The general definition of the soft function was given in Eq.~\eqref{S_def}.
For the $qq' \to qq'$ channel, the Wilson lines are
%---------------------------------------------------------------------------------------
\begin{align}
 \cW_{\bI} = \mathbf{T} 
    \left\{ 
        (Y^\dagger_\rF {\blue T_\bI} Y_\rT ) 
        (Y^\dagger_\rTH {\blue T_\bI} Y_\rO)  
    \right\}\,,
\end{align}
with ${\blue T_\bO} = {\blue \tau^a}$ and ${\blue T_\bT} = {\blue \mathbbm{1}}$. A
more explicit form of these operators, Eq.~\eqref{Sexplicit} is
%---------------------------------------------------------------------------------------
\begin{multline}
 S_{\bI \bJ} = 
  \sum_{X_s} \big\langle0\Big| \tmmathbf{\bar{T}}  
    \left\{  
      (Y_\rO^\dagger {\blue T^{\dagger}_I} Y_\rTH )^{\blue \ i_1}_{\blue i_3} 
      (Y_\rT^\dagger {\blue T^{\dagger}_I} Y_\rF )^{\blue \ i_2}_{\blue i_4} 
    \right\}  
    \Big| X_s \Big\rangle  \\
\times     \big\langle X_s\Big|  \tmmathbf{T}  
    \left\{  
      (Y_{\rF}^{\dagger} {\blue T_J} Y_{\rT} )^{\blue \ i_4}_{\blue i_2} 
      (Y_{\rTH}^{\dagger} {\blue T_J} Y_{\rO} )^{\blue \ i_3}_{\blue i_1} 
    \right\} \Big| 0 \Big\rangle \,
  \delta \!\left( {\red n_\rF} \cdt P^X_\rF - k_\rF \right)  
  \delta \! \left( {\red n_\rTH} \cdt P^X_\rTH - k_\rTH \right)  \,.
\end{multline}
%---------------------------------------------------------------------------------------
The tree-level color factor is
%---------------------------------------------------------------------------------------
\begin{equation}
   D_{\bI \bJ}^{\text{tree}}
= 
\begin{pmatrix}
  \frac{1}{2}C_F C_A & 0 \\
                   0 & C_A^2 \\
\end{pmatrix} \,,
\end{equation}
%---------------------------------------------------------------------------------------
and the 1-loop color factors are 
%---------------------------------------------------------------------------------------
\begin{align}
  D_{\bI \bJ} (\rO, \rT)
&= 
     D_{\bI \bJ}(\rTH, \rF)
    = 
    \frac{C_F}{2}
    \left(  \begin{array}{cc}
                1     & -C_A  \\
                - C_A &  0 
            \end{array}
    \right)\\
  D_{\bI \bJ}   (\rO,\rF)
&= 
    D_{\bI \bJ}   (\rT, \rTH)
    = 
    \frac{C_F}{4}
    \left(  \begin{array}{cc}
                (C_A^2 - 2) & \quad 2C_A \\
                2C_A        & 0          \\
            \end{array}
    \right)\\
   D_{\bI \bJ}  (\rO,\rTH)
&= 
     D_{\bI \bJ}  (\rT ,\rF)
    = 
    \frac{C_F}{4}
    \left(  \begin{array}{cc}
                -1 &  0        \\
                0             & 4C_A^2 \\
            \end{array}
    \right) \,.
\end{align}
%---------------------------------------------------------------------------------------
%---------------------------------------------------------------------------------------
%------------------------qq -> gg-------------------------------------------------------
%---------------------------------------------------------------------------------------
\subsubsection*{case
               $gg \to q\bar{q}$
               }
For the  $gg \to q\bar{q}$
channel,  the Wilson line structure is
%---------------------------------------------------------------------------------------
\begin{align}
\mathcal{W}^{ab}_\bI 
=
    Y_\rTH^\dagger 
    \left( 
      \mathcal{Y}_\rO^{T} 
      {\blue T}_\bI 
      \mathcal{Y}_\rT  
    \right)^{\ba \bb}  
    Y_\rF  
=  
    Y_\rTH^\dagger 
    \left( 
      \mathcal{Y}_\rO^{\ba \ba'} 
      {\blue T}_\bI^{\ba' \bb'} 
      \mathcal{Y}_\rT^{\bb' \bb}  
    \right)  
    Y_\rF  \,,
\end{align}
%---------------------------------------------------------------------------------------
where ${\blue T}_{\bI}^{\ba \bb}$ are 
%---------------------------------------------------------------------------------------
\begin{align}
{\blue T_{1}^{ab}} &= {\blue \tau^a \tau^b} \nn \\
{\blue T_{2}^{ab}} &= {\blue \tau^b \tau^a} \nn \\
{\blue T_{3}^{ab}} &= {\blue \delta^{ab} } \,,
\end{align}
%---------------------------------------------------------------------------------------
and $\mathcal{Y}_\rn^{\ba \bb}$ is a soft Wilson line in the adjoint representation.
The explicit form for the soft function for this channel is
%---------------------------------------------------------------------------------------
\begin{align}
 S_{\bI \bJ} &= 
  \sum_{X_s} \left\langle 0 \left|  \tmmathbf{\bar{T}}  
    \left\{  
        \left[
        Y_\rF^\dagger 
        \left( 
                \mathcal{Y}_\rT^{\dagger} 
                {\blue T}_{\bJ}^{\dagger}
                \mathcal{Y}_\rO
        \right)^{\bb \ba}  
        Y_\rTH  
        \right]^{\blue \ i_1}_{\blue i_2}
    \right\} \right| X_s \right\rangle
  \\
&~~~~~~~~
 \times 
\left\langle X_s \left|
 \tmmathbf{T}  
    \left\{
        \left[
        Y_\rTH^\dagger 
        \left( 
            \mathcal{Y}_\rO^{\dagger} 
            {\blue T_\bI}
            \mathcal{Y}_\rT
        \right)^{\ba \bb}   
        Y_\rF  
        \right]^{\blue \ i_2}_{\blue i_1}
    \right\} \right| 0 \right\rangle
\delta \left( {\red n_\rF} \cdt P^X_\rF - k_\rF \right)  \delta \left( {\red n_\rTH} \cdt P^X_\rTH - k_\rTH \right)
\nn
\,,
\end{align}
%---------------------------------------------------------------------------------------
where we have written the color indices explicitly, with $\ba,\bb$ being adjoint indices and 
${\blue i_1},{\blue i_2}$ being fundamental indices.  The tree level color factor is
%---------------------------------------------------------------------------------------
\begin{eqnarray}
D^{\text{tree}}_{\bI \bJ}
= 
\dfrac{C_F}{2}
\begin{pmatrix}
 2 C_A C_F & -1 & 2 C_A \\
 -1 & 2 C_A C_F & 2 C_A \\
 2 C_A & 2 C_A & 4 C_A^2
 \end{pmatrix} \,.
\end{eqnarray}
%---------------------------------------------------------------------------------------
The 1-loop color factors are
%---------------------------------------------------------------------------------------
\begin{align}
D_{\bI \bJ} (\rO,\rT)
    &= 
    \dfrac{C_F C_A^2}{4}
    \begin{pmatrix}
      C_A & 0 & 4 \\
      0 & C_A & 4 \\
      4 & 4 & 8 C_A
    \end{pmatrix}\\
D_{\bI \bJ} (\rTH, \rF)
    &= 
    \frac{C_F}{4C_A}
    \left(
    \begin{array}{ccc}
        1 & C_A^2+1 & 4 C_A^2 C_F \\
        C_A^2+1 & 1 & 4 C_A^2 C_F \\
        4 C_A^2 C_F & 4 C_A^2 C_F & 8 C_A^3 C_F
    \end{array}
    \right)
\\ 
D_{\bI \bJ} (\rO, \rTH)
&= 
D_{\bI \bJ} (\rT, \rF)
    = 
    \dfrac{C_F C_A}{4}
    \begin{pmatrix}
      2 C_A C_F & -1     & 2 C_A \\
      -1        & -1     & -2 C_A \\
      2 C_A     & -2 C_A & 0
    \end{pmatrix}\\
D_{\bI \bJ} (\rO,\rF)
&= 
D_{\bI \bJ} (\rT, \rTH)
    = 
    \dfrac{C_F C_A}{4}
    \begin{pmatrix}
      -1 & -1 & -2 C_A \\
      -1 & 2 C_A C_F & 2 C_A \\
      -2 C_A & 2 C_A & 0
    \end{pmatrix} \,.
\end{align}
%---------------------------------------------------------------------------------------
%---------------------------------------------------------------------------------------
%------------------------------------   gggg--------------------------------------------
%---------------------------------------------------------------------------------------
\subsubsection*{case:   $ gg \to gg $ }
The Wilson line structure for $gg \to gg$ is
%---------------------------------------------------------------------------------------
\begin{eqnarray}
\mathcal{W}^{\ba \bb \bc \bd}_{\bI} 
&=&  
    {\blue T}^{\ba' \bb' \bc' \bd'}_{\bI}
    \mathcal{Y}_\rO^{\ba \ba'} 
    \mathcal{Y}_\rT^{\bb \bb'} 
    \mathcal{Y}_\rTH^{\bc \bc'} 
    \mathcal{Y}_\rF^{\bd \bd'} \,,
\end{eqnarray}
%---------------------------------------------------------------------------------------
where the ${\blue T}^{\ba \bb \bc \bd}_{\bI}$ are given by
%---------------------------------------------------------------------------------------
\begin{align}
{\blue T^{a b c d}_{1} } &= \text{Tr}[{\blue \tau^{a} \tau^{b}\tau^{c}\tau^{d} } ] &  
{\blue T^{a b c d}_{6} } &= \text{Tr}[{\blue \tau^{a} \tau^{c}\tau^{b}\tau^{d} } ] \nn \\ 
{\blue T^{a b c d}_{2} } &= \text{Tr}[{\blue \tau^{a} \tau^{b}\tau^{d}\tau^{c} } ] & 
{\blue T^{a b c d}_{7} } &= \text{Tr}[{\blue \tau^{a} \tau^{d} }] \text{Tr}[{\blue \tau^{c}\tau^{b} } ]\nn \\ 
{\blue T^{a b c d}_{3} } &= \text{Tr}[{\blue \tau^{a} \tau^{d}\tau^{c}\tau^{b} } ] &
{\blue T^{a b c d}_{8} } &= \text{Tr}[{\blue \tau^{a} \tau^{b} }] \text{Tr}[{\blue \tau^{c}\tau^{d} } ]\nn \\ 
{\blue T^{a b c d}_{4} } &= \text{Tr}[{\blue \tau^{a} \tau^{d}\tau^{b}\tau^{c} } ] & 
{\blue T^{a b c d}_{9} } &= \text{Tr}[{\blue \tau^{a} \tau^{c} }] \text{Tr}[{\blue \tau^{b}\tau^{b} } ]\nn \,. \\ 
{\blue T^{a b c d}_{5} } &= \text{Tr}[{\blue \tau^{a} \tau^{c}\tau^{d}\tau^{b} } ] 
\label{glue:color}
\end{align}
%---------------------------------------------------------------------------------------
The soft function for this channel is
%---------------------------------------------------------------------------------------
\begin{align}
 S_{\bI \bJ} &= 
  \sum_{X_s} \bra{0}  \tmmathbf{\bar{T}}  
    \left\{  
         ( \mathcal{W}_\bJ^{\blue a b c d} )^{\dagger}
    \right\}  \ket{X_s}  \nonumber \bra{X_s}  \tmmathbf{T}  
    \left\{
        \mathcal{W}_\bI^{\blue a b c d }
    \right\} \ket{0} 
% \nonumber \\&\times 
\delta \left( {\red n_\rF} \cdt P^X_\rF - k_\rF \right)  \delta \left( {\red n_\rTH} \cdt P^X_\rTH - k_\rTH \right)
\,.
\end{align}
%---------------------------------------------------------------------------------------
The tree level color factor for $N=3$ is
%---------------------------------------------------------------------------------------
\begin{eqnarray}
    D^{\text{tree}}_{\bI \bJ}
 = 
\frac{1}{6} \left(
\begin{array}{ccccccccc}
 4 & -2 & 19 & -2 & -2 & -2 & 8 & 8 & -1 \\
 -2 & 4 & -2 & -2 & 19 & -2 & -1 & 8 & 8 \\
 19 & -2 & 4 & -2 & -2 & -2 & 8 & 8 & -1 \\
 -2 & -2 & -2 & 4 & -2 & 19 & 8 & -1 & 8 \\
 -2 & 19 & -2 & -2 & 4 & -2 & -1 & 8 & 8 \\
 -2 & -2 & -2 & 19 & -2 & 4 & 8 & -1 & 8 \\
 8 & -1 & 8 & 8 & -1 & 8 & 24 & 3 & 3 \\
 8 & 8 & 8 & -1 & 8 & -1 & 3 & 24 & 3 \\
 -1 & 8 & -1 & 8 & 8 & 8 & 3 & 3 & 24
\end{array}
\right)\,.
\end{eqnarray}
%---------------------------------------------------------------------------------------
The 1-loop results, also with $N = 3$, are
%---------------------------------------------------------------------------------------
\begin{align}
     D_{\bI \bJ}(\rO,\rT)
    &= 
     D_{\bI \bJ}(\rTH,\rF)
    = 
    \frac{1}{12}
    \left(
    \begin{array}{ccccccccc}
     20 & 2 & 65 & -7 & 2 & -7 & 27 & 48 & 0 \\
     2 & 20 & 2 & -7 & 65 & -7 & 0 & 48 & 27 \\
     65 & 2 & 20 & -7 & 2 & -7 & 27 & 48 & 0 \\
     -7 & -7 & -7 & -16 & -7 & -16 & -27 & -6 & -27 \\
     2 & 65 & 2 & -7 & 20 & -7 & 0 & 48 & 27 \\
     -7 & -7 & -7 & -16 & -7 & -16 & -27 & -6 & -27 \\
     27 & 0 & 27 & -27 & 0 & -27 & 0 & 18 & -18 \\
     48 & 48 & 48 & -6 & 48 & -6 & 18 & 144 & 18 \\
     0 & 27 & 0 & -27 & 27 & -27 & -18 & 18 & 0
    \end{array}
    \right)
  \\
     D_{\bI \bJ}(\rO,\rTH)
    &= 
     D_{\bI \bJ}(\rT, \rF)
    = 
    \frac{1}{12}
    \left(
    \begin{array}{ccccccccc}
     -16 & -7 & -16 & -7 & -7 & -7 & -27 & -27 & -6 \\
     -7 & 20 & -7 & 2 & 65 & 2 & 0 & 27 & 48 \\
     -16 & -7 & -16 & -7 & -7 & -7 & -27 & -27 & -6 \\
     -7 & 2 & -7 & 20 & 2 & 65 & 27 & 0 & 48 \\
     -7 & 65 & -7 & 2 & 20 & 2 & 0 & 27 & 48 \\
     -7 & 2 & -7 & 65 & 2 & 20 & 27 & 0 & 48 \\
     -27 & 0 & -27 & 27 & 0 & 27 & 0 & -18 & 18 \\
     -27 & 27 & -27 & 0 & 27 & 0 & -18 & 0 & 18 \\
     -6 & 48 & -6 & 48 & 48 & 48 & 18 & 18 & 144
    \end{array}
    \right)
    \\
     D_{\bI \bJ}(\rO,\rF)
    &= 
     D_{\bI \bJ}(\rT,\rTH)
    = 
    \frac{1}{12}
    \left(
    \begin{array}{ccccccccc}
     20 & -7 & 65 & 2 & -7 & 2 & 48 & 27 & 0 \\
     -7 & -16 & -7 & -7 & -16 & -7 & -6 & -27 & -27 \\
     65 & -7 & 20 & 2 & -7 & 2 & 48 & 27 & 0 \\
     2 & -7 & 2 & 20 & -7 & 65 & 48 & 0 & 27 \\
     -7 & -16 & -7 & -7 & -16 & -7 & -6 & -27 & -27 \\
     2 & -7 & 2 & 65 & -7 & 20 & 48 & 0 & 27 \\
     48 & -6 & 48 & 48 & -6 & 48 & 144 & 18 & 18 \\
     27 & -27 & 27 & 0 & -27 & 0 & 18 & 0 & -18 \\
     0 & -27 & 0 & 27 & -27 & 27 & 18 & -18 & 0
    \end{array}
    \right)\,.
\end{align}
%---------------------------------------------------------------------------------------

\subsection{Final threshold thrust soft function}
With the integrals done and the color factors calculated, all that remains is to combine them together to form the
threshold thrust soft function.
 The threshold thrust soft function is determined from the hemisphere soft function by
%---------------------------------------------------------------------------------------
\begin{align}
  S_{\bI \bJ}(k) &= \int\rd k_\rTH \rd k_\rF S_{\bI \bJ}(k_\rTH,k_\rF) \delta(k-k_\rTH-k_\rF)\\
&=
\delta(k) D^{\text{tree}}_{\bI \bJ} 
+\left(\frac{\alpha_s}{4\pi}\right)
 \frac{\mu^{2\e}}{k^{1 + 2 \varepsilon}} \sum_{{\red a},{\red b}}
\left[\phantom{\frac{1}{6}}\!\!\!
R_{\red ab}(\rnu )+
L_{\red ab}(\rnu )
\right] 
 D_{\bI \bJ}(\chi(\ra),\chi( \rb)) \,.
\end{align}
%---------------------------------------------------------------------------------------
This is straightforward to compute from the results in this Appendix, so we will just give a few representative examples. To compute
the regulated soft function, we expand in $\e$ using
\begin{equation} \label{distexp}
\frac{\mu^{2\e}}{k^{1+2\e}} = -\frac{1}{2\varepsilon}\delta(k) + \left[ \frac{1}{k} \right]_\star^{[k,\mu]}
-2\varepsilon \left[ \frac{1}{k} \ln
\frac{k}{\mu} \right]_\star^{[k,\mu]}+\cdots \,,
\end{equation}
and drop the $\frac{1}{\e}$ and $\frac{1}{\e^2}$ terms using $\overline{\text{MS}}$ subtraction. The $[]_\star^{[k,\mu]}$ is
a star distribution (see for example~\cite{Schwartz:2007ib} for a definition and some relations).

In the $qq' \to  qq'$ channel ($stu$), for which $\chi(\ra) = \ra$, the threshold thrust soft function is
%---------------------------------------------------------------------------------------
\begin{multline}
S_{\bI \bJ}(k) = \delta(k) 
  \begin{pmatrix}
            \frac{1}{2}C_F C_A      &  0  \\
                      0             &  C_A^2
  \end{pmatrix} 
+\left(\frac{\alpha_s}{4\pi}\right)  c^S_{\bI \bJ}(\rnu)\delta(k)\\
+\left(\frac{\alpha_s}{4\pi}\right)   
\begin{pmatrix}
      4 C_F \ln\frac{(1+\rnu)^2}{\rnu} - 8 C_F^2 C_A \ln\rnu\,\Theta(1-\rnu) &
     -8C_A C_F \ln (1 + \rnu) \\
     -8C_A C_F \ln (1 + \rnu) &
      16 C_A^2 C_F \ln \rnu\, \Theta(\rnu-1)
\end{pmatrix}
\left[\frac{1}{k}\right]_\star^{ [k,\mu]} \,.
\end{multline}
%---------------------------------------------------------------------------------------
The constant ($\mu$-independent) part,  $c^S_{\bI \bJ}(\rnu)$ are messy functions, 
which take different form for $\rnu>1$ and $\rnu <1$. For example, the $11$ component 
of the $qq' \to qq'$ threshold thrust
soft function is
%---------------------------------------------------------------------------------------
\begin{multline}
c^S_{\blue 11}(\rnu)
=C_F \Big[\left(C_A^2+2\right) 
\LI\left(\frac{1}{\rnu}\right)
+\left(C_A^2-2\right) \Big(-\ln ^2(\rnu -1)+\ln \frac{\rnu-1}{\rnu}\Big)+4 \LI(1-\rnu )\\
+\left(1-3 C_A^2\right) \ln ^2\rnu+ \left(4 C_A^2-6\right) \ln\rnu \ln (\rnu -1) 
+4 \ln^2(\rnu +1)\\-4 \ln\rnu \ln (\rnu +1)+\pi^2\Big]\Theta(\rnu-1)\\
+C_F\Big[- \left(2 C_A^2+1\right) \Big( \LI(1-\rnu ) + \ln \rnu \ln (1-\rnu)\Big)
+ \left(C_A^2-4\right)\ln^2\rnu+ 4 \ln^2(\rnu +1)\\
-4 \ln (\rnu +1) \ln \rnu- \ln(1-\rnu)
+\frac{\pi^2}{6} \left(2C_A^2+7\right)\\
+4 \LI\left(\frac{\rnu -1}{\rnu }\right)+ \ln ^2(1-\rnu )\Big]\Theta(1-\rnu) \label{csqqqq}
\,.
\end{multline}
%---------------------------------------------------------------------------------------
The other constants have similarly complicated expressions and we
see no reason to write them out explicitly.

To compute the soft function for a crossing, one sums the integrals against the color
factors with the Wilson lines permuted as in Tables~\ref{tab:perm1} and~\ref{tab:perm2}. For example, 
the soft function for $q \bar{q} \to q' \bar{q}'$ ($ust$) comes out as
%---------------------------------------------------------------------------------------
\begin{multline}
S_{\bI \bJ}(k) = \delta(k) \label{line1}
\left( \begin{smallmatrix}
            \frac{1}{2}C_F C_A      &  0  \\
                      0             &  C_A^2
 \end{smallmatrix}\right)
+\left(\frac{\alpha_s}{4\pi}\right)
c^S_{\bI \bJ}(\rnu)\delta(k) \\
+\left(\frac{\alpha_s}{4\pi}\right)
\left(\begin{smallmatrix}
    -4  C_F \ln[ (1+\rnu)\rnu]+ 8 C_F^2 C_A \ln\rnu\,\Theta(\rnu-1) &
    \quad  8C_A C_F \ln \rnu \\
    \quad  8C_A C_F \ln \rnu &
    \quad   16C_A^2 C_F \Big[\ln (1 + \rnu) - \ln \rnu\, \Theta(1-\rnu)\Big]
 \end{smallmatrix}\right)
\left[\frac{1}{k}\right]_\star^{ [k,\mu]} 
\,.
\end{multline}
%---------------------------------------------------------------------------------------
These $c^S_{\bI \bJ}(\rnu)$ are different from the  $c^S_{\bI \bJ}(\rnu)$ for $q q' \to q q'$,
but not worth writing out explicitly.

In the $gg \to q\bar{q}$ channel,
%---------------------------------------------------------------------------------------
\begin{multline}
S_{\bI \bJ}(k) = 
\left(
  \begin{smallmatrix}
    C_A C_F^2      & -\frac{C_F}{2} & C_A C_F \\
    -\frac{C_F}{2} & C_A C_F^2      & C_A C_F \\
    C_A C_F        & C_A C_F        & 2 C_A^2 C_F
  \end{smallmatrix}
\right)
\delta(k)
+ \left(\frac{\alpha_s}{4\pi}\right)
\left\{
 c^S_{\bI \bJ}( \rnu) \delta(k)
{\phantom{
  \begin{smallmatrix}
    C_A C_F^2      & -\frac{C_F}{2} & C_A C_F \\
    -\frac{C_F}{2} & C_A C_F^2      & C_A C_F \\
    C_A C_F        & C_A C_F        & 2 C_A^2 C_F%
  \end{smallmatrix}
}}
\right.
\\
+
\left(
 \begin{smallmatrix}
   C_A C_F^2      & -\frac{C_F}{2} & C_A C_F \\
    -\frac{C_F}{2} & C_A C_F^2      & C_A C_F \\
    C_A C_F        & C_A C_F        & 2 C_A^2 C_F
  \end{smallmatrix}
\right)
\left(
16(C_A-C_F)
\left[\frac{  \ln{ \frac{k}{\mu}} }{k} \right]_\star^{[k,\mu]}
 -16C_A \ln \rnu\Theta(1-\rnu)
\left[\frac{1}{k} \right]_\star^{[k,\mu]}
\right)
\\
\left.
+
4C_F C_A
\left(
 \begin{smallmatrix}
    2 C_F C_A \ln \rnu + C_A^2 \ln(1+\rnu)                    & 
    - \ln\rnu &
    4 C_A \ln (1+\rnu) + 2 C_A \ln \rnu      \\
    - \ln \rnu & 
    C_A^2 \ln (1+\rnu)- \ln \rnu                      & 
    -2C_A \ln \rnu + 4 C_A\ln (1+\rnu)                     \\
   2C_A \ln \rnu +4 C_A \ln (1+\rnu)        & 
    -2C_A \ln \rnu + 4C_A\ln (1+\rnu)                    & 
    8 C_A^2  \ln (1+\rnu)
 \end{smallmatrix}
\right)
\left[\frac{1}{k} \right]_\star^{[k,\mu]}
\right\}
\,.
\end{multline}
%---------------------------------------------------------------------------------------
The functions $c^S_{\bI \bJ}(\rnu)$ are again different from the other channels, and to messy
to write out. Note the additional $[\ln(k)/k]_\star^{[k,\mu]}$ factor in this channel,
which happened to be absent for the four quark channels.


\begin{thebibliography}{99}

\bibitem{GehrmannDeRidder:2007hr}
  A.~Gehrmann-De Ridder, T.~Gehrmann, E.~W.~N.~Glover and G.~Heinrich,
  %``NNLO corrections to event shapes in $e^+e^-$ annihilation,''
  JHEP {\bf 0712}, 094 (2007)
  [arXiv:0711.4711 [hep-ph]].
  %%CITATION = JHEPA,0712,094;%%

\bibitem{Becher:2008cf}
  T.~Becher and M.~D.~Schwartz,
  %``A Precise determination of $\alpha_s$ from LEP thrust data using effective
  %field theory,''
  JHEP {\bf 0807}, 034 (2008)
  [arXiv:0803.0342 [hep-ph]].
  %%CITATION = JHEPA,0807,034;%%

\bibitem{Fleming:2007qr}
  S.~Fleming, A.~H.~Hoang, S.~Mantry and I.~W.~Stewart,
  %``Jets from massive unstable particles: Top-mass determination,''
  Phys.\ Rev.\  D {\bf 77}, 074010 (2008)
  [arXiv:hep-ph/0703207].
  %%CITATION = PHRVA,D77,074010;%%

\bibitem{Schwartz:2007ib}
  M.~D.~Schwartz,
  %``Resummation and NLO Matching of Event Shapes with Effective Field Theory,''
  Phys.\ Rev.\  D {\bf 77}, 014026 (2008)
  [arXiv:0709.2709 [hep-ph]].
  %%CITATION = PHRVA,D77,014026;%%

\bibitem{Abbate:2010xh}
  R.~Abbate, M.~Fickinger, A.~H.~Hoang, V.~Mateu and I.~W.~Stewart,
  %``Thrust at N^3LL with Power Corrections and a Precision Global Fit for
  %alphas(mZ),''
  arXiv:1006.3080 [hep-ph].
  %%CITATION = ARXIV:1006.3080;%%

\bibitem{Kaplan:2008pt}
  D.~E.~Kaplan and M.~D.~Schwartz,
  %``Constraining Light Colored Particles with Event Shapes,''
  Phys.\ Rev.\ Lett.\  {\bf 101}, 022002 (2008)
  [arXiv:0804.2477 [hep-ph]].
  %%CITATION = PRLTA,101,022002;%%

\bibitem{Dasgupta:2001sh}
  M.~Dasgupta and G.~P.~Salam,
  %``Resummation of non-global QCD observables,''
  Phys.\ Lett.\  B {\bf 512}, 323 (2001)
  [arXiv:hep-ph/0104277].
  %%CITATION = PHLTA,B512,323;%%

\bibitem{Dasgupta:2002dc}
  M.~Dasgupta and G.~P.~Salam,
  %``Resummed event-shape variables in DIS,''
  JHEP {\bf 0208}, 032 (2002)
  [arXiv:hep-ph/0208073].
  %%CITATION = JHEPA,0208,032;%%

\bibitem{Becher:2007ty}
  T.~Becher, M.~Neubert and G.~Xu,
  %``Dynamical Threshold Enhancement and Resummation in Drell-Yan Production,''
  JHEP {\bf 0807}, 030 (2008)
  [arXiv:0710.0680 [hep-ph]].
  %%CITATION = JHEPA,0807,030;%%

\bibitem{Becher:2009th}
  T.~Becher and M.~D.~Schwartz,
  %``Direct photon production with effective field theory,''
  JHEP {\bf 1002}, 040 (2010)
  [arXiv:0911.0681 [hep-ph]].
  %%CITATION = JHEPA,1002,040;%%

\bibitem{Ahrens:2010zv}
  V.~Ahrens, A.~Ferroglia, M.~Neubert, B.~D.~Pecjak and L.~L.~Yang,
  %``Renormalization-Group Improved Predictions for Top-Quark Pair Production at
  %Hadron Colliders,''
  arXiv:1003.5827 [hep-ph].
  %%CITATION = ARXIV:1003.5827;%%

\bibitem{Laenen:1998qw}
  E.~Laenen, G.~Oderda and G.~F.~Sterman,
  %``Resummation of threshold corrections for single particle inclusive
  %cross-sections,''
  Phys.\ Lett.\  B {\bf 438}, 173 (1998)
  [arXiv:hep-ph/9806467].
  %%CITATION = PHLTA,B438,173;%%


%\cite{Chiu:2008vv}
\bibitem{Chiu:2008vv}
  J.~y.~Chiu, R.~Kelley and A.~V.~Manohar,
  %``Electroweak Corrections using Effective Field Theory: Applications to the
  %LHC,''
  Phys.\ Rev.\  D {\bf 78}, 073006 (2008)
  [arXiv:0806.1240 [hep-ph]].
  %%CITATION = PHRVA,D78,073006;%%


\bibitem{Chiu:2009ft}
  J.~y.~Chiu, A.~Fuhrer, R.~Kelley and A.~V.~Manohar,
  %``Soft and Collinear Functions for the Standard Model,''
  Phys.\ Rev.\  D {\bf 81}, 014023 (2010)
  [arXiv:0909.0947 [hep-ph]].
  %%CITATION = PHRVA,D81,014023;%%

\bibitem{Chiu:2009mg}
  J.~y.~Chiu, A.~Fuhrer, R.~Kelley and A.~V.~Manohar,
  %``Factorization Structure of Gauge Theory Amplitudes and Application to Hard
  %Scattering Processes at the LHC,''
  Phys.\ Rev.\  D {\bf 80}, 094013 (2009)
  [arXiv:0909.0012 [hep-ph]].
  %%CITATION = PHRVA,D80,094013;%%

\bibitem{Chiu:2009yz}
  J.~y.~Chiu, A.~Fuhrer, A.~H.~Hoang, R.~Kelley and A.~V.~Manohar,
  %``Using SCET to calculate electroweak corrections in gauge boson
  %production,''
  PoS E {\bf FT09}, 009 (2009)
  [arXiv:0905.1141 [hep-ph]].
  %%CITATION = POSCI,EFT09,009;%%


\bibitem{Becher:2009cu}
  T.~Becher and M.~Neubert,
  %``Infrared singularities of scattering amplitudes in perturbative QCD,''
  Phys.\ Rev.\ Lett.\  {\bf 102}, 162001 (2009)
  [arXiv:0901.0722 [hep-ph]].
  %%CITATION = PRLTA,102,162001;%%

\bibitem{Becher:2009qa}
  T.~Becher and M.~Neubert,
  %``On the Structure of Infrared Singularities of Gauge-Theory Amplitudes,''
  JHEP {\bf 0906}, 081 (2009)
  [arXiv:0903.1126 [hep-ph]].
  %%CITATION = JHEPA,0906,081;%%

\bibitem{Kelley:2010fn}
  R.~Kelley and M.~D.~Schwartz,
  %``1-loop matching and NNLL resummation for all partonic 2 to 2 processes in
  %QCD,''
  arXiv:1008.2759 [hep-ph].
  %%CITATION = ARXIV:1008.2759;%%

\bibitem{Fuhrer:2010eu}
  A.~Fuhrer, A.~V.~Manohar, J.~y.~Chiu and R.~Kelley,
  %``Radiative Corrections to Longitudinal and Transverse Gauge Boson and Higgs
  %Production,''
  Phys.\ Rev.\  D {\bf 81}, 093005 (2010)
  [arXiv:1003.0025 [hep-ph]].
  %%CITATION = PHRVA,D81,093005;%%

\bibitem{Kidonakis:1998nf}
  N.~Kidonakis, G.~Oderda and G.~F.~Sterman,
  %``Evolution of color exchange in {QCD} hard scattering,''
  Nucl.\ Phys.\  B {\bf 531}, 365 (1998)
  [arXiv:hep-ph/9803241].
  %%CITATION = NUPHA,B531,365;%%

\bibitem{Botts:1989kf}
  J.~Botts and G.~F.~Sterman,
  %``Hard Elastic Scattering In QCD: Leading Behavior,''
  Nucl.\ Phys.\  B {\bf 325}, 62 (1989).
  %%CITATION = NUPHA,B325,62;%%

\bibitem{Kunszt:1993sd}
  Z.~Kunszt, A.~Signer and Z.~Trocsanyi,
  %``One Loop Helicity Amplitudes For All 2 $\to$ 2 Processes In QCD And N=1
  %Supersymmetric Yang-Mills Theory,''
  Nucl.\ Phys.\  B {\bf 411}, 397 (1994)
  [arXiv:hep-ph/9305239].
  %%CITATION = NUPHA,B411,397;%%

\bibitem{Bern:1990cu}
  Z.~Bern and D.~A.~Kosower,
  %``Efficient calculation of one loop QCD amplitudes,''
  Phys.\ Rev.\ Lett.\  {\bf 66}, 1669 (1991).
  %%CITATION = PRLTA,66,1669;%%

\bibitem{Chien:2010kc}
  Y.~T.~Chien and M.~D.~Schwartz,
  %``Resummation of heavy jet mass and comparison to LEP data,''
  JHEP {\bf 1008}, 058 (2010)
  [arXiv:1005.1644 [hep-ph]].
  %%CITATION = JHEPA,1008,058;%%

\bibitem{Banfi:2010xy}
  A.~Banfi, G.~P.~Salam and G.~Zanderighi,
  %``Phenomenology of event shapes at hadron colliders,''
  JHEP {\bf 1006}, 038 (2010)
  [arXiv:1001.4082 [hep-ph]].
  %%CITATION = JHEPA,1006,038;%%

\bibitem{Banfi:2004yd}
  A.~Banfi, G.~P.~Salam and G.~Zanderighi,
  %``Principles of general final-state resummation and automated
  %implementation,''
  JHEP {\bf 0503}, 073 (2005)
  [arXiv:hep-ph/0407286].
  %%CITATION = JHEPA,0503,073;%%

\bibitem{Banfi:2004nk}
  A.~Banfi, G.~P.~Salam and G.~Zanderighi,
  %``Resummed event shapes at hadron - hadron colliders,''
  JHEP {\bf 0408}, 062 (2004)
  [arXiv:hep-ph/0407287].
  %%CITATION = JHEPA,0408,062;%%

\bibitem{Stewart:2009yx}
  I.~W.~Stewart, F.~J.~Tackmann and W.~J.~Waalewijn,
  %``Factorization at the LHC: From PDFs to Initial State Jets,''
  Phys.\ Rev.\  D {\bf 81}, 094035 (2010)
  [arXiv:0910.0467 [hep-ph]].
  %%CITATION = PHRVA,D81,094035;%%

\bibitem{Becher:2006mr}
  T.~Becher, M.~Neubert and B.~D.~Pecjak,
  %``Factorization and momentum-space resummation in deep-inelastic
  %scattering,''
  JHEP {\bf 0701}, 076 (2007)
  [arXiv:hep-ph/0607228].
  %%CITATION = JHEPA,0701,076;%%

\bibitem{Bauer:2003pi}
  C.~W.~Bauer and A.~V.~Manohar,
  %``Shape function effects in B --> X/s gamma and B --> X/u l nu decays,''
  Phys.\ Rev.\  D {\bf 70}, 034024 (2004)
  [arXiv:hep-ph/0312109].
  %%CITATION = PHRVA,D70,034024;%%

\bibitem{Manohar:2003vb}
  A.~V.~Manohar,
  %``Deep inelastic scattering as x --> 1 using soft-collinear effective
  %theory,''
  Phys.\ Rev.\  D {\bf 68}, 114019 (2003)
  [arXiv:hep-ph/0309176].
  %%CITATION = PHRVA,D68,114019;%%

\bibitem{Becher:2006nr}
  T.~Becher and M.~Neubert,
  %``Threshold resummation in momentum space from effective field theory,''
  Phys.\ Rev.\ Lett.\  {\bf 97}, 082001 (2006)
  [arXiv:hep-ph/0605050].
  %%CITATION = PRLTA,97,082001;%%

\bibitem{Becher:2010pd}
  T.~Becher and G.~Bell,
  %``The gluon jet function at two-loop order,''
  arXiv:1008.1936 [hep-ph].
  %%CITATION = ARXIV:1008.1936;%%

\bibitem{Gordon:1993qc}
  L.~E.~Gordon and W.~Vogelsang,
  %``Polarized and unpolarized prompt photon production beyond the leading
  %order,''
  Phys.\ Rev.\  D {\bf 48}, 3136 (1993).
  %%CITATION = PHRVA,D48,3136;%%

\bibitem{Fleming:2007xt}
  S.~Fleming, A.~H.~Hoang, S.~Mantry and I.~W.~Stewart,
  %``Top Jets in the Peak Region: Factorization Analysis with NLL Resummation,''
  Phys.\ Rev.\  D {\bf 77}, 114003 (2008)
  [arXiv:0711.2079 [hep-ph]].
  %%CITATION = PHRVA,D77,114003;%%

\bibitem{Ellis:2010rw}
  S.~D.~Ellis, C.~K.~Vermilion, J.~R.~Walsh, A.~Hornig and C.~Lee,
  %``Jet Shapes and Jet Algorithms in SCET,''
  arXiv:1001.0014 [hep-ph].
  %%CITATION = ARXIV:1001.0014;%%

\bibitem{Bauer:2010vu}
  C.~W.~Bauer, N.~D.~Dunn and A.~Hornig,
  %``Factorization of Boosted Multijet Processes for Threshold Resummation,''
  arXiv:1002.1307 [hep-ph].
  %%CITATION = ARXIV:1002.1307;%%

\bibitem{deFlorian:2007fv}
  D.~de Florian and W.~Vogelsang,
  %``Resummed cross-section for jet production at hadron colliders,''
  Phys.\ Rev.\  D {\bf 76}, 074031 (2007)
  [arXiv:0704.1677 [hep-ph]].
  %%CITATION = PHRVA,D76,074031;%%

\bibitem{Stewart:2010tn}
  I.~W.~Stewart, F.~J.~Tackmann and W.~J.~Waalewijn,
  %``N-Jettiness: An Inclusive Event Shape to Veto Jets,''
  arXiv:1004.2489 [hep-ph].
  %%CITATION = ARXIV:1004.2489;%%

\bibitem{Cheung:2009sg}
  W.~Y.~Cheung, M.~Luke and S.~Zuberi,
  %``Phase Space and Jet Definitions in SCET,''
  Phys.\ Rev.\  D {\bf 80}, 114021 (2009)
  [arXiv:0910.2479 [hep-ph]].
  %%CITATION = PHRVA,D80,114021;%%

\bibitem{Forshaw:2006fk}
  J.~R.~Forshaw, A.~Kyrieleis and M.~H.~Seymour,
  %``Super-leading logarithms in non-global observables in QCD,''
  JHEP {\bf 0608}, 059 (2006)
  [arXiv:hep-ph/0604094].
  %%CITATION = JHEPA,0608,059;%%


\end{thebibliography}
\end{document}